\documentclass[journal=tosc]{iacrtrans}

\usepackage{graphicx}
\usepackage{subfigure}
\usepackage{booktabs}
\usepackage{makecell}
\usepackage[ruled,linesnumbered]{algorithm2e}
\usepackage{threeparttable}
\usepackage{appendix}
\usepackage{float}
\usepackage{multirow}
\usepackage{bm}
\usepackage{amsmath}
\usepackage{amssymb}
\usepackage{color}
\usepackage{mathrsfs}
\usepackage[misc,geometry]{ifsym}
\usepackage[bookmarks=true, bookmarksopen=true]{hyperref}
\newcommand{\simon}{\textsf{Simon}\xspace}
\newcommand{\simeck}{\textsf{Simeck}\xspace}
\author{Siwei Chen\inst{1} \and Zejun Xiang$^*$\inst{1} \and Xiangyong Zeng\inst{2} \and Guangxue Qin\inst{1}}
\institute{
	School of Cyber Science and Technology, Key Laboratory of Intelligent Sensing System \\and Security, Hubei University, Wuhan, China
	\and
	Faculty of Mathematics and Statistics, Hubei Key Laboratory of Applied Mathematics,\\Hubei University, Wuhan, China\\
	\email{chensiwei\_hubu@163.com, {xiangzejun, xzeng}@hubu.edu.cn}
}

\title{Enhancing the MILP/MIQCP-based Automatic Search for Differential-Linear Distinguishers of Simon-Like Ciphers}

\begin{document}
	
	\maketitle


	\begin{abstract}
			In this paper, we propose an improved method based on Mixed-Integer Linear Programming/Mixed-Integer Quadratic Constraint Programming (MILP/MIQCP) to automatically find better differential-linear (DL) distinguishers for the all members of \simon and \simeck block cipher families. To be specific, we first give the completely precise MILP model to describe the linear part, and explain how to utilize the general expressions of \textsf{Gurobi} solver to model the propagation of continuous difference for the middle part in a quite easy way. Secondly, in order to solve the MILP/MIQCP model in a reasonable time, we propose two heuristic strategies based on the divide-and-conquer idea to speed up the search process. Thirdly, we introduce the transforming technique, which exploits the clustering effect on DL trails, to improve the estimated correlation of the DL approximation. 
			
			We apply our method to Simon and Simeck block cipher families. Consequently, we find the 14/17/21/26-round theoretical DL distinguishers of \simon32/48/64/96, which extend the previous longest ones of \simon32/48/96 by one round and \simon64 by two rounds, respectively. For \simeck, we do not explore longer distinguishers compared to the currently best results, but refresh all the results of Zhou \textit{et al}. (the first work to automate finding DL distinguishers for \simon-like ciphers using MILP/MIQCP). Besides, in order to validate the correctness of these distinguishers, the experimental verifications are conducted on \simon32/\simeck32 and \simon48/\simeck48. The results show that our theoretical estimations on correlations are very close to the experimental ones, which can be regarded as a concrete support for the effectiveness of our method.

	\end{abstract}
	
	\section{Introduction}
With the rapid development of the Internet of Things (IoT) technology, some wireless and portable devices such as smart band, smart phone, unmanned aerial vehicle etc, which can achieve an immediate interaction with human, are more and more widely applied in daily life. The ciphers used in these small and light devices require not only the strong security but also the low-latency and low-energy. Therefore, the design as well as the cryptanalysis on the lightweight symmetric-key primitives have been the very hot topics in cryptography. In the past two decades, a number of lightweight primitives are presented successively such as \textsf{PRESENT}~\cite{present}, \textsf{CLEFIA}~\cite{clefia}, 
\textsf{Trivium}~\cite{trivium}, 
\textsf{Grain}~\cite{grain}, 
 \textsf{MICKEY}~\cite{mickey}, \textsf{KATAN} and \textsf{KTANTAN}~\cite{katan},
\textsf{Quark}~\cite{quark}, \textsf{Piccolo}~\cite{piccolo},  \textsf{PHOTON}~\cite{photon}, \textsf{SPONGENT}~\cite{spongent}, \textsf{PRINCE}~\cite{prince}, \textsf{Simon} and \textsf{Speck}~\cite{DBLP:journals/iacr/BeaulieuSSTWW13}, \textsf{RECTANGLE}~\cite{rectangle},  \textsf{Simeck}~\cite{DBLP:conf/ches/YangZSAG15}, \textsf{Midori}~\cite{midori}, \textsf{SKINNY}~\cite{skinny}, 
\textsf{Gimli}~\cite{gimli},
 \textsf{LIZARD}~\cite{lizard}, \textsf{Orthros}~\cite{orthros},
\textsf{ASCON}~\cite{ascon} etc. Also, these primitives have attracted a lot of cryptographers to analysis the security weakness. In general, when designing a new symmetric primitive, the designers always consider its security margin against the currently common cryptanalysis methods to ensure that it cannot be attacked completely at least for the foreseeable future.

Nowadays, differential and linear cryptanalysis are the two of the most effective methods and essential citizens for evaluating the security of symmetric-key primitives, especially for the block ciphers. At CRYPTO '90, Biham and Shamir~\cite{DBLP:conf/crypto/BihamS90} gave the concept of differential cryptanalysis and succeeded in the theoretical attack of the full-round Data Encryption Standard (\textsf{DES}) at the first time. Later, Mutsui~\cite{DBLP:conf/eurocrypt/Matsui93} proposed linear cryptanalysis at EUROCRYPT '93 to improve the full-round attack on \textsf{DES}. Since their proposal, some derivative variants based on them were put forward like truncated differential cryptanalysis~\cite{truncated-diff}, higher-order differential cryptanalysis~\cite{truncated-diff,Lai-higher-order-diff}, differential-linear crytpanalysis~\cite{DBLP:conf/crypto/LangfordH94}, etc. Compared to the original differential or linear cryptanalysis, these variant methods may not be widely applicable to various block ciphers but can achieve the better attacks for some certain ciphers. For example, the best record on the key-recovery attacks of \textsf{Serpent} is kept by~\cite{crypto/LiuLL21} using differential-linear cryptanalysis.

Differential-linear (DL) cryptanalysis was first introduced by Langford and Hellman~\cite{DBLP:conf/crypto/LangfordH94} at CRYPTO '94 to launch an 8-round key-recovery attack on \textsf{DES}. First, they split a target cipher $E$ into two independent sub-ciphers: the differential part $E_d$ and the linear part $E_l$ as $E=E_l\circ E_d$. Then, they prepared a (truncated) differential with probability 1 for $E_d$ and a linear approximation with correlation $q$ for $E_l$, where the active bits of $E_l$'s input mask only appear at the inactive bit positions of $E_d$'s output difference. Finally, a DL approximation with correlation $q^2$ for $E$ was derived by merging the two sub-ciphers. Afterwards, at ASIACRYPT '02, Biham \textit{et al}.~\cite{DBLP:conf/asiacrypt/BihamDK02} generalized DL cryptanalysis by seeking the (truncated) differential with probability $p$ ($p < 1$) for $E_d$ and re-evaluated the correlation as $pq^2$ under the assumption that $E_d$ is completely independent with $E_l$. However, this assumption hardly holds in practice and the experimental evaluation on the correlation is always far from the predicted one. For instance, Dobrauning \textit{et al.}~\cite{DBLP:conf/ctrsa/DobraunigEMS15} presented a DL distinguisher for 4-round \textsf{ASCON} permutation with a predicted correlation of $2^{-19}$ whereas the test one is $2^{-1}$. To handle this problem, at EUROCRYPT '19, Bar-On \textit{et al.}~\cite{DBLP:conf/eurocrypt/Bar-OnDKW19} divided $E$ into three parts as $E = E_l \circ E_m \circ E_d$ instead of the previous two parts and gave the concept of differential-linear connectivity table (DLCT), which is inspired by the boomerang connectivity table (BCT)~\cite{DBLP:conf/eurocrypt/CidHPSS18}. Since then, a general approach to explore DL distinguishers is to search for best differential (resp. linear approximation) for $E_d$ (resp. $E_l$) using automatic tools, then evaluate the correlation of $E_m$ by experiments. Nevertheless, how to transform the computation on the correlation of $E_m$ to an automatic solving problem and further achieve a fully automatic DL cryptanalysis were still unsolved until last year. 

At CT-RSA 2023, Bellini \textit{et al.}~\cite{DBLP:conf/ctrsa/BelliniGGMP23} proposed a fully automated DL cryptanalysis for ARX ciphers. They adopted the three-part DL cryptanalysis structure and modeled the correlation of $E_m$ by characterizing the continuous difference propagation of the modular addition, rotation and XOR operations with MILP/MIQCP. By this, they explored the best DL distinguishers for \textsf{Speck32/64}. Later, for \textsf{Simon}-like ciphers, Zhou \textit{et al.}~\cite{journals/iet/ZhouWH24} investigated the continuous difference propagation of the round function and achieved the fully automated DL distinguishers search with MILP/MIQCP. Also, they explored the best DL distinguishers for the specific versions of \textsf{Simon} and \textsf{Simeck}. Afterwords, Hadipour \textit{et al.}~\cite{DBLP:journals/iacr/HadipourDE24} proposed an automatic tool based on CP/MILP to search for DL distinguishers. They also adopted the Bar-On \textit{et al.}'s three-part structure but focused on extending the generalized DLCT (i.e., the middle part) via a boomerang perspective. By applying this tool to various symmetric-key primitives, they either presented the first DL distinguishers or enhanced the best ones including \textsf{Simeck}. Very recently, Li \textit{et al.}~\cite{journals/iet/LiZH24} proposed a method called $\mathcal{K}6$ to search for DL distinguishers for AND-RX ciphers. Different from the previous methods, they used a truncated differential and a linear hull to construct the DL distinguisher. They applied $\mathcal{K}6$ to \textsf{Simon} and \textsf{Simeck}, then found the best DL distinguisher of \textsf{Simon128}. 

Nevertheless, as regards the DL distinguishers search for \textsf{Simon}-like ciphers in~\cite{journals/iet/ZhouWH24}, there are several problems that are worth further studying. Firstly, when constructing the model for the linear part $E_l$, they just regarded the bit-wise AND operation as the independent S-boxes, which is not precise enough. Secondly, the obtained DL distinguisher by the automatic method is actually the optimal single trail, similarly to the single differential characteristic or linear trail in classical differential or linear cryptanalysis. Thus, whether the estimated correlation can be further improved or not if taking the clustering effect into account is unknown. In fact, this problem also exists in Bellini \textit{et al.}'s work~\cite{DBLP:conf/ctrsa/BelliniGGMP23}.
At last, Zhou \textit{et al.} reused the piece-wise linear function, which was given in~\cite{DBLP:conf/ctrsa/BelliniGGMP23}, to approximate the logarithm on the absolute value of a general variable when modeling the middle part, and implemented the process by MiniZinc. It indeed works but may be not friendly or easy-to-use for users. Motivated by these problems, especially the second one, we will give a deep study on searching for DL distinguishers of \textsf{Simon}-like ciphers in this paper.
	\subsection{Our Contributions}
	In this paper, we follow Zhou \textit{et al.}'s work~\cite{journals/iet/ZhouWH24} and aim to improve the MILP/MIQCP-based automatic method to find better DL distinguishers for \textsf{Simon}-like ciphers. Our contributions consist of the following four parts.
	
		\vspace{5pt}
		\noindent \textbf{Constructing MILP/MIQCP model with precision and ease.} An entire model to search for DL distinguishers is composed of three sub-models that describe the propagation of the differential, the continuous difference and the linear approximation, respectively. In this paper, we model the linear approximation propagation based on K\"{o}lbl \textit{et al.}'work~\cite{DBLP:conf/crypto/KolblLT15}, which clarifies the exact linear behavior of the round function of \textsf{Simon}-like ciphers. Compared to regarding the bit-wise AND operation as a series of two-to-one independent Sboxes in~\cite{journals/iet/ZhouWH24}, our model is completely precise. In addition, we simplify the modeling for the logarithm on absolute value of a general variable using the \textit{general expressions} `\texttt{abs\_()}' and `\texttt{addGenConstrLogA()}', which are \textsf{Python} interfaces provided by \textsf{Gurobi}\footnote{\textsf{Gurobi} optimization. \url{https://www.gurobi.com/}} optimizer. Compared to the piece-wise approximation function presented in~\cite{DBLP:conf/ctrsa/BelliniGGMP23}, our model looks more intuitive and the modeling process is much easier to implement.
		
		\vspace{5pt}
		\noindent\textbf{Speeding up the search process with heuristic strategies.} Note that the sub-model for the middle part is composed of quadratic constraints and general variables, thus the entire model that generated by merging the three sub-models is too complex to be solved by \textsf{Gurobi} in a reasonable time.  
		However, it will take a long time to solve even a single model when the round or block size is large. To tackle this problem, we propose two heuristic strategies, called differential-first and linear-first strategies, based on the divide-and-conquer idea. Both of them aim to reduce the solution space by dividing the entire model to two parts, and then speed up the search process by solving the two sub-models separately. For a given round configuration, our strategies may not explore the best DL approximation, but can find a good one efficiently. For example, it takes about 24 hours to find a DL distinguisher with correlation $2^{-19.49}$ for the 15-round \textsf{Simon48} using configuration $(5,5,5)$. However, it only takes no more than 40 seconds to get a better DL distinguisher with correlation $2^{-18.66}$ with the differential-first strategy.
		
		\vspace{5pt}
		\noindent\textbf{Taking the differential-linear clustering effect into account.} In~\cite{DBLP:conf/ctrsa/BelliniGGMP23,journals/iet/ZhouWH24}, there always exist the significant gaps between the estimated and the experimental correlations of the presented DL distinguishers. A potential reason that cannot be ignored is that the estimated correlation is actually the one of trails (we call DL trail later) contained within the corresponding DL distinguisher, which is similar to the best single differential/linear characteristic.
		Previous works~\cite{DBLP:conf/crypto/KolblLT15,DBLP:conf/asiacrypt/LeurentPS21} have shown the strong clustering effect on the differential/linear characteristic in \textsf{Simon}-like ciphers. Naturally, if we also take the clustering effect on the DL approximation into account, the estimated correlation of a DL distinguisher is possibly more close to the exact one. To achieve this goal, we present a trivial but effective approach, named \textit{transforming technique}, to compute and sum over the correlations of all possible DL trails that share the common difference-mask pair. By this technique, a DL distinguisher with a high correlation can be derived from a given DL trail. For example, we re-evaluate the 13-round DL distinguisher of \textsf{Simon32} presented by~\cite{journals/iet/ZhouWH24} with the transforming technique. Consequently, its estimated correlation is improved from $2^{-16.63}$ to $2^{-13.94}$, which is very close to the test correlation $2^{-13.19}$. 
		
		\vspace{5pt}
		\noindent\textbf{Presenting the best DL distinguishers for \textsf{Simon} ciphers.} We apply the enhanced method to searching for DL distinguishers for \textsf{Simon} and \textsf{Simeck}. As a consequence, we present the better DL distinguishers for the all versions of \textsf{Simon} family compared to the previous work in terms of either the round or the correlation. Especially, we extend the previous best theoretical DL distinguishers of $\textsf{Simon32/48/96}$ by one round and $\textsf{Simon64}$ by two rounds, and improve the estimated correlation of 32-round DL distinguisher of \textsf{Simon128}. Regarding \textsf{Simeck}, our results are weaker than the currently best ones~\cite{DBLP:journals/iacr/HadipourDE24}, but the improvements compared to the all DL distinguishers presented by~\cite{journals/iet/ZhouWH24} are very significant. Our results and the comparison with the previous works are summarized in Table~\ref{tab:result-summary}. In addition, we verify the newly found and re-evaluated DL distinguishers of \textsf{Simon32}/\textsf{Simeck32} and \textsf{Simon48}/\textsf{Simeck48} by experiments. The result (see in Table~\ref{tab:experiments} of Sect.~\ref{Sect:experiments}) shows that our estimated correlations are very close to the test ones, which also indicates that our improved method is actually effective. 
		

	
\begin{table}[!h]
	\caption{DL distinguishers of \textsf{Simon} and \textsf{Simeck}. \#R: the number of round. \textsf{Cor}: the correlation. Symbol `$\star$': invalid distinguisher in theory due to its data complexity $\frac{\epsilon}{\textsf{Cor}^{2}}$, where $\epsilon$ is set to 1 in general.}\label{tab:result-summary}
	\centering  
	\subtable{  
		\begin{tabular}{clcc}
			\toprule
			Cipher&\#R&$\mathsf{Cor}$&Ref.\\
			\midrule
			\multirow{7}{*}{\textsf{Simon32}}&11&$2^{-13.89}$&\cite{journals/iet/LiZH24}\\
			&\textbf{11}&$\bm{2^{-8.03}}$&Sect.~\ref{Sect:4.2}\\
			&$13^{\star}$&$2^{-16.63}$&\cite{journals/iet/ZhouWH24}\\
			&\textbf{13}&$\bm{2^{-11.99}}$ &Sect.~\ref{Sect:4.2}\\
			&$14^{\star}$&$2^{-18.63}$ &\cite{journals/iet/ZhouWH24}\\
			&\textbf{14}&$\bm{2^{-15.36}}$ &Sect.~\ref{Sect:4.2}\\
			&$\textbf{15}^{\star}$&$\bm{2^{-18.29}}$ &Sect.~\ref{Sect:4.2}\\
			\midrule
			\multirow{9}{*}{\textsf{Simon48}}
			&14&$2^{-21.30}$ & \cite{journals/iet/LiZH24}\\
			&\textbf{14}&$\bm{2^{-13.33}}$&Sect.~\ref{Sect:4.2}\\
			&15&$2^{-20.19}$ &\cite{journals/iet/ZhouWH24}\\
			&\textbf{15}&$\bm{2^{-15.70}}$ &Sect.~\ref{Sect:4.2}\\
			&16&$2^{-22.66}$&\cite{journals/iet/ZhouWH24}\\
			&\textbf{16}&$\bm{2^{-18.90}}$&Sect.~\ref{Sect:4.2}\\
			&$17^{\star}$&$2^{-24.66}$&\cite{journals/iet/ZhouWH24}\\
			&\textbf{17}&$\bm{2^{-21.88}}$&Sect.~\ref{Sect:4.2}\\
			&$\textbf{18}^{\star}$&$\bm{2^{-24.55}}$&Sect.~\ref{Sect:4.2}\\
			\midrule
			\multirow{5}{*}{\textsf{Simon64}}&16&$2^{-23.31}$&\cite{journals/iet/LiZH24}\\
			&\textbf{16}&$\bm{2^{-16.26}}$&Sect.~\ref{Sect:4.2}\\
			&$20^{\star}$&$2^{-34.58}$&\cite{journals/iet/ZhouWH24}\\
			&\textbf{20}&$\bm{2^{-28.67}}$&Sect.~\ref{Sect:4.2}\\
			&\textbf{21}&$\bm{2^{-31.99}}$&Sect.~\ref{Sect:4.2}\\
			\midrule
			\multirow{6}{*}{\textsf{Simon96}}
			&23&$2^{-46.57}$&\cite{journals/iet/LiZH24}\\
			&\textbf{23}&$\bm{2^{-31.52}}$&Sect.~\ref{Sect:4.2}\\
			&25&$2^{-46.66}$&\cite{journals/iet/ZhouWH24}\\
			&\textbf{25}&$\bm{2^{-40.85}}$&Sect.~\ref{Sect:4.2}\\
			&$26^{\star}$&$2^{-50.66}$&\cite{journals/iet/ZhouWH24}\\
			&\textbf{26}&$\bm{2^{-44.03}}$&Sect.~\ref{Sect:4.2}\\

			\bottomrule
		\end{tabular} 
		\label{tab:firsttable}  
	}  
	\subtable{          
	\begin{tabular}{clcc}
		\toprule
		Cipher&\#R&\textsf{Cor}&Ref.\\
		
		\midrule
		\multirow{5}{*}{\textsf{Simon128}}
		&31&$2^{-62.70}$&\cite{journals/iet/ZhouWH24}\\
		&\textbf{31}&$\bm{2^{-55.66}}$&Sect.~\ref{Sect:4.2}\\
		&$32^{\star}$&$2^{-66.70}$&\cite{journals/iet/ZhouWH24}\\
		&32&$2^{-59.75}$&\cite{journals/iet/LiZH24}\\
		
		&\textbf{32}&$\bm{2^{-59.62}}$&Sect.~\ref{Sect:4.2}\\
		\midrule
		\multirow{5}{*}{\textsf{Simeck32}}&12&$2^{-13.89}$ &\cite{journals/iet/LiZH24}\\
		&\textbf{12}&$\bm{2^{-9.00}}$&Sect.~\ref{Sect:4.2}\\
		&$14^{\star}$&$2^{-16.63}$ &\cite{journals/iet/ZhouWH24}\\
		&\textbf{14}&$\bm{2^{-14.73}}$ &Sect.~\ref{Sect:4.2}\\
		&14&$2^{-13.92}$ &\cite{DBLP:journals/iacr/HadipourDE24}\\
		\midrule
		\multirow{7}{*}{\textsf{Simeck48}}&17&$2^{-22.37}$ &\cite{journals/iet/ZhouWH24}\\
		&17&$2^{-21.54}$ &\cite{journals/iet/LiZH24}\\
		&\textbf{17}&$\bm{2^{-17.79}}$ &Sect.~\ref{Sect:4.2}\\
		&$18^{\star}$&$2^{-24.75}$&\cite{journals/iet/ZhouWH24}\\
		&\textbf{18}&$\bm{2^{-20.41}}$&Sect.~\ref{Sect:4.2}\\
		&\textbf{19}&$\bm{2^{-23.22}}$&Sect.~\ref{Sect:4.2}\\
		&20&$2^{-21.89}$&\cite{DBLP:journals/iacr/HadipourDE24}\\
		\midrule
		\multirow{10}{*}{\textsf{Simeck64}}&22&$2^{-32.44}$&\cite{journals/iet/ZhouWH24}\\
		&22&$2^{-30.41}$&\cite{journals/iet/LiZH24}\\
		&\textbf{22}&$\bm{2^{-25.98}}$&Sect.~\ref{Sect:4.2}\\
		&$23^{\star}$&$2^{-36.13}$&\cite{journals/iet/ZhouWH24}\\
		&\textbf{23}&$\bm{2^{-30.92}}$&Sect.~\ref{Sect:4.2}\\
		&$24^{\star}$&$2^{-38.13}$&\cite{journals/iet/ZhouWH24}\\
		&\textbf{24}&$\bm{2^{-31.88}}$&Sect.~\ref{Sect:4.2}\\
		&$25^{\star}$&$2^{-41.04}$&\cite{journals/iet/ZhouWH24}\\
		&$\textbf{25}^{\star}$&$\bm{2^{-35.53}}$&Sect.~\ref{Sect:4.2}\\
		&26&$2^{-30.35}$&\cite{DBLP:journals/iacr/HadipourDE24}\\
		\bottomrule
	\end{tabular} 
	\label{tab:secondtable}  
	}  
\end{table}	
	\subsection{Organization of This Paper}
	The rest of this paper is organized as follows. Section~\ref{Sect:2} provides some notations throughout this paper, a brief description on the \textsf{Simon}-like ciphers, and the introduction to differential-linear cryptanalysis, continuous difference and MILP/MIQCP. In Section~\ref{Sect:3}, we exhibit how to construct the model for finding DL trails in a more precise and easier way, and introduce two heuristic strategies to speed up the search process as well as the transforming technique to obtain a real DL distinguisher. In Section~\ref{Sect:4}, we apply the improved method to \textsf{Simon}-like ciphers and list our superior results. Finally, we conclude this paper in Section~\ref{Sect:5}.


	\section{Preliminaries}\label{Sect:2}
	In this section, we start by introducing some notations throughout this paper in Table~\ref{notaion}, and then give a brief introduction to the \textsf{Simon}-like ciphers, differential-linear cryptanalysis, continuous difference and MILP/MIQCP.
	\begin{table}[!h]
		\centering
		\caption{Some notations of this paper. Apart from the follows, we denote the $i$th coordinate of an $n$-dimensional vector $x=(x_0,x_1,...,x_{n-1})$ by $x_i$.}{\label{notaion}}
		\begin{tabular}{cl}
			\toprule
			Notation & Description\\
			\midrule
			$\mathbb{F}_2$ & A finite field only contains two elements, i.e. $\{0, 1\}$ \\
			$\mathbb{F}_2^{n}$ & An $n$-dimensional vectorial space defined over $\mathbb{F}_2$ \\
			$\mathbb{R}$ & A field contains all the real numbers\\
			$\mathbb{Z}$ & A set contains all the integers\\
			$\vee, \wedge, \oplus$ & Bit-wise OR, AND, XOR \\
			$S^{t}(x)$& Circular left shift of $x$ by $t$ bits\\
			$S^{-t}(x)$ & Circular right shift of $x$ by $t$ bits\\
			$\overline{x}$ & Bit-wise negation\\
			$wt(x)$ & Hamming weight of $x$\\
			$x||y$ & Concatenation of vectors $x$ and $y$\\
			$0^{n}, 1^{n}$& The vectors of $\mathbb{F}_{2}^{n}$ with all 0s and all 1s\\
			\bottomrule
		\end{tabular}
		
	\end{table}

	\subsection{Description of \textsf{Simon}-like Ciphers}
\begin{figure}[!htb]
		\centering
		\includegraphics[width=0.5\textwidth]{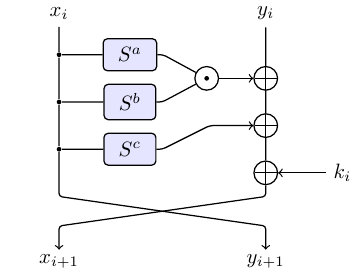}
		\caption{Round function of \textsf{Simon}-like ciphers.}\label{fig:simon-like}
	\end{figure}

	The \textsf{Simon}-like ciphers mainly include two families of lightweight block ciphers: \textsf{Simon} and \textsf{Simeck}. They are designed based on the classical Feistel structure and adopt the similar round function, which only consists of three types of bit-wise operations: AND ($ \wedge $), XOR ($ \oplus $) and cyclic rotation by $\lambda$ bits (i.e., $S^{t}$). The $i$th round encryption, as depicted in Figure~\ref{fig:simon-like}, can be represented by 
	\begin{equation*}
		(x_{i+1},y_{i+1}) = (S^{a}(x_i)\wedge S^{b}(x_i)\oplus S^{c}(x_i)\oplus y_i \oplus k_i, x_i),
	\end{equation*}
	where $k_i$ denotes the $i$th round key. 
	The differences between \textsf{Simon} and \textsf{Simeck} are mainly reflected on the rotation offsets $(a,b,c)$, block size, key size and key schedules. \textsf{Simon}~\cite{DBLP:journals/iacr/BeaulieuSSTWW13} was presented by the National Security Agency (NSA) in 2013. There are total ten members of the family commonly denoted by \textsf{Simon}2$\textit{n/mn}$, where the block size is 2\textit{n} for $n \in\{16, 24, 32, 48, 64\}$ and the key size is $\textit{mn}$ for $m \in\{2, 3, 4\}$. All the versions use the rotation offset $(8,1,2)$ in round function and utilize linear feed back register to generate roundkeys. Later, Yang et al.~\cite{DBLP:conf/ches/YangZSAG15} chose $(5,0,1)$ as the rotation offset and proposed \textsf{Simeck} block ciphers at CHES 2015, to pursue a better performance in hardware implementation. There are three variants of \textsf{Simeck} denoted by \textsf{Simeck}2$\textit{n}$ for $ n \in \{16,24,32\}$ where the block size is $2n$ and the key size is $4n$. In addition, \textsf{Simeck} reuses its round function as the key schedule to generate roundkeys. In this paper, we focus on searching for DL disinguishers in the single-key setting, thus we omit the introduction on the key schedules of \textsf{Simon} and \textsf{Simeck}. Please refer to~\cite{DBLP:journals/iacr/BeaulieuSSTWW13} and~\cite{DBLP:conf/ches/YangZSAG15} for more details about the specifications on the two cipher families.

	\subsection{Differential-Linear Cryptanalysis}\label{Sect:2.2}
\begin{figure}[h]
		\centering
		\subfigure[Two-part structure.]{
			\label{fig:two-part-DL} 
			\includegraphics[scale=0.4]{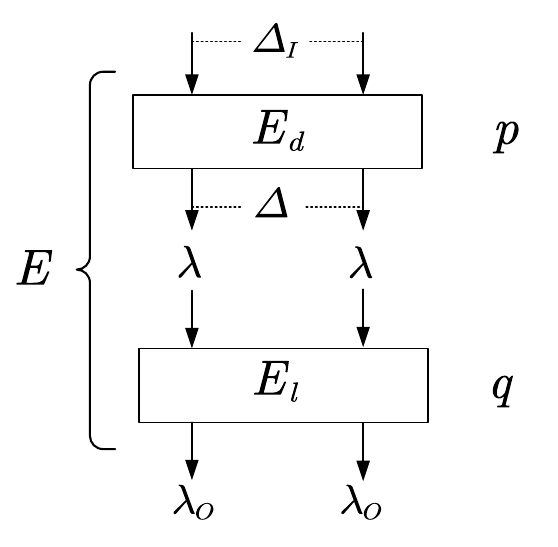}}
		\hspace{0.5in} 
		\subfigure[Three-part structure.]{
			\label{fig:three-part-DL} 
			\includegraphics[scale=0.4]{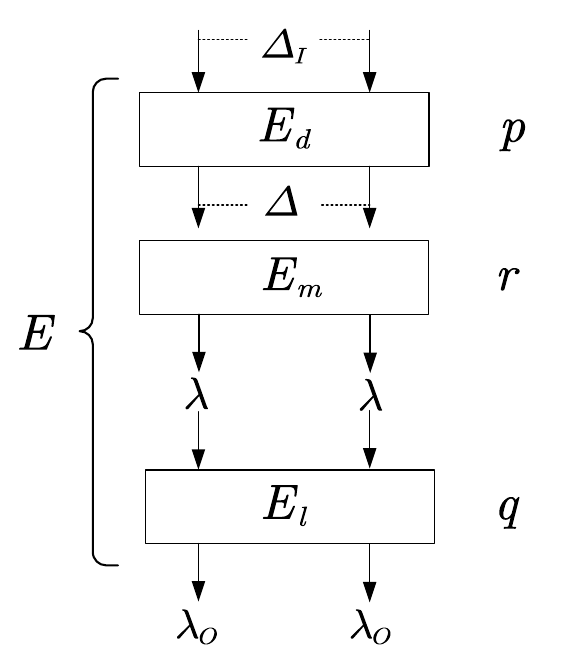}}
		\caption{Two structures of DL distinguishers.}
		\label{fig1}
	\end{figure}

	Differential-linear cryptanalysis was originally introduced by Langford and Hellman~\cite{DBLP:conf/crypto/LangfordH94} at CRYPTO '94, and was first generalized by Biham \textit{et al.}~\cite{DBLP:conf/asiacrypt/BihamDK02} at ASIACRYPT '02. The Figure~\ref{fig:two-part-DL} shows the generalized structure of a DL distinguisher, of which the correlation is evaluated as $pq^2$ under the assumption that $E_d$ is completely independent with $E_l$ where $p$ (resp. $q$) is the differential probability (resp. linear approximation correlation). Because the independence assumption hardly holds in practice, so Bar-On \textit{et al.}~\cite{DBLP:conf/eurocrypt/Bar-OnDKW19} introduced the DLCT and gave a new structure of three parts, as shown in Figure~\ref{fig:three-part-DL}. Combining this new structure with the exact evaluation on correlation of DL distinguishers~\cite{DBLP:journals/joc/BlondeauLN17}, they gave the following formula to evaluate the correlation of three-part DL distinguisher:
	\begin{equation}\label{equ:DLCT-formula}
		\textsf{Cor}_E(\Delta_I,\lambda_O) =\sum_{\Delta,\lambda}\textsf{Pr}_{E_d}(\Delta_I,\Delta)\cdot\textsf{Cor}_{E_m}(\Delta,\lambda)\cdot\textsf{Cor}^2_{E_l}(\lambda,\lambda_O),
	\end{equation}
	where $\textsf{Pr}_{E_d}(\Delta_I,\Delta)$ denotes the probability of differential $\Delta_I\overset{E_d}{\rightarrow}\Delta$, $\textsf{Cor}_{E_m}(\Delta,\lambda)$ denotes the correlation of DL approximation $\Delta\overset{E_m}{\rightarrow}\lambda$, and $\textsf{Cor}_{E_l}(\lambda,\lambda_O)$  denotes the correlation of linear approximation $\lambda\overset{E_l}{\rightarrow}\lambda_O$.
	
	For a given differential-mask pair $(\Delta_I,\lambda_O)$ of $E$ with block size $n$, Equation~\eqref{equ:DLCT-formula} is precise enough to evaluate the correlation but quite difficult to practically implement since it requires to enumerate the all difference-mask pairs $(\Delta, \lambda)$ of $E_m$. Therefore, the common approach to explore a DL distinguisher based on the three-part structure in practice is to find a good differential $\Delta_I\overset{E_d}{\rightarrow}\Delta$ with probability $p$ and a good linear approximation $\lambda\overset{E_f}{\rightarrow}\lambda_O$ with correlation $q$, then experimentally compute the DL approximation $\Delta\overset{E_m}{\rightarrow}\lambda$ with correlation $r$, and finally estimate this DL distinguisher with correlation $prq^2$. In fact, this approach considers only one DL trail and regards its correlation as that of the DL distinguisher. Towards the automatic method~\cite{DBLP:conf/ctrsa/BelliniGGMP23,journals/iet/ZhouWH24}, strictly, it just explores the single DL trail with the optimal combination $(\Delta_I,\Delta,\lambda,\lambda_O)$ among the all such that the correlation is the best, instead of finding the real distinguisher, which can be represented as
	\begin{equation}\label{equ:DL-trail}
		|\textsf{Cor}^*_E|=\max_{(\Delta_I,\Delta,\lambda,\lambda_O)}\left\{\textsf{Pr}_{E_d}(\Delta_I,\Delta)\cdot|\textsf{Cor}_{E_m}(\Delta,\lambda)|\cdot\textsf{Cor}^2_{E_l}(\lambda,\lambda_O)\right\}.
	\end{equation}
	To facilitate the distinction and avoid the confusion, we call the DL approximation obtained respectively by Equation~\eqref{equ:DLCT-formula} and Equation~\eqref{equ:DL-trail} as the DL distinguisher and DL trail in Section~\ref{Sect:3} and~\ref{Sect:4}.
	
	\subsection{The Continuous Difference and Its Propagation Rules}
	The concept of \textit{continuous difference} was firstly introduced by Bellini \textit{et al.}~\cite{DBLP:conf/ctrsa/BelliniGGMP23}, which is inspired by the continuous generalization of various cryptographic operations~\cite{DBLP:journals/access/CoutinhoJB20}. In fact, it expresses the differential behavior of one bit using the correlation instead of the probability. For example, suppose that the difference value of a bit equaling to $0$ holds with probability $p$, then its continuous difference\footnote{In order to keep consistent with the meaning of correlation in DL/linear cryptanalysis, we represent the continuous difference of a bit throughout this paper as $2p-1$ whereas it is actually $1-2p$ in~\cite{DBLP:conf/ctrsa/BelliniGGMP23}, where $p$ is the probability that the difference of this bit equals to 0. But it does not matter because we only focus on the absolute value.} can be calculated as $2p-1$. Apparently, the correlation of a DL distinguisher can be regarded as the continuous difference of the linear combination on the output bits. So, if we can describe the propagation of continuous differences for the underlying cryptographic operations contained in a cipher, then finding a good DL distinguisher will become much easier. The following propositions give the propagation rules of continuous difference for XOR, left rotation and AND operations.
	\begin{proposition}[Continuous difference propagation of XOR~\cite{DBLP:conf/ctrsa/BelliniGGMP23}]\label{prop:CF-XOR-rule}
	For the XOR operation $z = x\oplus y$, where $x,y,z$ belong to $\mathbb{F}_2$ and $x$ is independent with y. Assume that the continuous differences corresponding to the three bits are $\mathcal{C}_x,\mathcal{C}_y$ and $\mathcal{C}_z$, then we have 
	\[
		\mathcal{C}_z = \mathcal{C}_x \cdot\mathcal{C}_y.
	\]  
	\end{proposition}	
	\begin{proposition}[Continuous difference propagation of left rotation~\cite{DBLP:conf/ctrsa/BelliniGGMP23}]\label{prop:CF-rotation-rule}
		For the left operation $y = S^{m}(x)$, where $x,y$ belong to $\mathbb{F}_2^n$. Assume that the continuous difference corresponding to the $i$th bit of $x_i$ are $\mathcal{C}_{x_i}$, then the continuous difference of $y_i$ for $i\in\{0,1,...,n-1\}$ can be expressed as 
		\[
		\mathcal{C}_{y_i} = \mathcal{C}_{x_j}
		\]  
		where $j=(i+m)\mod n$.
	\end{proposition}	
	\begin{proposition}[Continuous difference propagation of AND~\cite{journals/iet/ZhouWH24}]\label{prop:CF-AND-rule}
	For the AND operation $z = x\wedge y$, where $x,y,z$ belong to $\mathbb{F}_2$ and $x$ is independent with $y$. Assume that the continuous differences corresponding to the three bits are $\mathcal{C}_x,\mathcal{C}_y$ and $\mathcal{C}_z$, then we have 
	\[
	\mathcal{C}_z = \frac{1}{4}(\mathcal{C}_x \cdot \mathcal{C}_y+\mathcal{C}_x+\mathcal{C}_y+1).
	\]  
	\end{proposition}

	\subsection{MILP and MIQCP}
	The Mixed-Integer Linear Programming (MILP) method is widely applied to solving the optimization problem. In general, it needs to construct a model that consists of a system of purely linear constraints, a series of integer and general variables and an linear objective function. By the opening solvers such as \textsf{Gurobi}, \textsf{CPLEX}, \textsf{SCIP} etc, the MILP model can be solved and the optimal solution is accessible. The Mixed-Integer Quadratic Constraint Programming (MIQCP) is a generalization of MILP. The most significant difference is that the MIQCP model contains the quadratic constraints even thought the objective function with the algebraic degree of 2.
	
	The first application of MILP to cryptanalysis for symmetric-key cryptosystem was introduced by Mouha \textit{et al.}~\cite{DBLP:conf/cisc/MouhaWGP11} at Inscrypt 2011. Since then, the automated cryptanalysis based on MILP has been a hot topic. Also, more and more automatic methods based on MILP as well as the modeling techniques are proposed successively including but not limited to~\cite{DBLP:conf/asiacrypt/SunHWQMS14,DBLP:conf/asiacrypt/XiangZBL16,DBLP:conf/crypto/TodoIHM17,DBLP:journals/tosc/BouraC20,DBLP:journals/tosc/LiS22}.

	\section{Finding DL Distinguishers with MILP/MIQCP Method}\label{Sect:3} 
	In this section, we first optimize the MILP/MIQCP model for finding the best DL trail by improving the precision of linear part sub-model and simplifying the constraint of middle part sub-model. To speed up the search process of DL trails, we then propose two heuristic strategies based on the divide-and-conquer idea, which enable us to find a good DL trail in a reasonable time. Finally, we introduce a trivial but effective technique to transform a DL trail to a DL distinguisher by exploring the valuable DL trails.
	
	\subsection{Optimizing the MILP/MIQCP Model}\label{Sect:3.1}
	To generate a model to find the best DL trail, we need to construct three sub-models separately for differential, middle and linear parts, and then add some constraints to merge them to an entire one. For the target cipher with $n$-bit branch and the rotation offset $(a,b,c)$, we next give the specific constraints and the corresponding algorithms for each part. 
	
	\paragraph{Modeling the differential part.} At CRYPTO 2015, K\"{o}lbl \textit{et al.}~\cite{DBLP:conf/crypto/KolblLT15} investigated the exact differential (resp. linear) behavior for \textsf{Simon}-like ciphers in theory and proposed the SMT/SAT based method to search for differential (resp. linear) characteristic. The differential propagation through the round function of \textsf{Simon}-like ciphers is illustrated as Theorem~\ref{thm:simon_pro}.
	\begin{theorem}[\cite{DBLP:conf/crypto/KolblLT15}]\label{thm:simon_pro}
		Let $f(x)=S^{a}(x)\wedge S^{b}(x)\oplus S^{c}(x)$, where $gcd(n,a-b)=1$, n is even, and $a>b$. Let $\alpha$ and $\beta$ be the input and output difference of $f(x)$. Let\\
		\begin{equation*}
			\mathtt{varibits}=S^{a}(\alpha)\vee S^{b}(\alpha)
		\end{equation*}
		and
		\begin{equation*}
			\mathtt{doublebits}=S^{b}(\alpha)\wedge \overline{S^{a}(\alpha)}\wedge S^{2a-b}(\alpha)
		\end{equation*}
		and
		\begin{equation*}
			\gamma=\beta\oplus S^{c}(\alpha).
		\end{equation*}
		We have that the probability that difference $\alpha$ goes to difference $\beta$ is
		\begin{equation*}
			\Pr(\alpha\xrightarrow{f}\beta)=\begin{cases}
				2^{-n+1} &\text{if $\alpha=1^n$ and $wt(\gamma)\equiv 0\mod2$},\\
				2^{-wt(\mathtt{varibits}\oplus \mathtt{doublebits})} &\text{if $\alpha\neq1^n$ and $\gamma\wedge(\overline{\mathtt{varibits}})=0^n$}\\
				&\text{and $(\gamma\oplus S^{a-b}(\gamma))\wedge \mathtt{doublebits}=0^n$},\\
				0 & \text{otherwise.}
			\end{cases}
		\end{equation*}
	\end{theorem}
In~\cite{journals/iet/ZhouWH24}, a system of linear inequalities is given to constrain the differential propagation through round function due to Theorem~\ref{thm:simon_pro}. 
Here we reuse their constraints as detailed in Appendix~\ref{appendix:A} and give the procedure as illustrated in Algorithm~\ref{alg:model-diff-part}, to construct an MILP model to describe the $R_d$-round differential part. 

\paragraph{Modeling the middle part with ease.} 
For the middle part, suppose that the internal state bits are independent of each other. Then we need to describe the propagation of continuous difference with MIQCP based on Proposition~\ref{prop:CF-XOR-rule},~\ref{prop:CF-rotation-rule} and~\ref{prop:CF-AND-rule}. The details of modeling the $r$-round propagation is given as below.

We use two vectorial variables $x^{r+1}$ and $x^{r}$, where they both belong to $\mathbb{R}^n$, to denote the left and right branches of $r$-round input continuous difference in the middle part, respectively. Also, we introduce an auxiliary vectorial variable $y^r\in\mathbb{R}^n$ to represent the continuous difference after $S^a(x^{r+1})$ and $S^b(x^{r+1})$ through the AND operation. According to the propagation rule illustrated in Proposition~\ref{prop:CF-AND-rule}, the relation between $x^{r+1}$ and $y^r$ can be described by the following quadratic constraint:
\begin{equation}\label{equ:model_AND-CD}
	4y^r_i = 1 + x^{r+1}_{i+a} + x^{r+1}_{i+b} + x^{r+1}_{i+a}\cdot x^{r+1}_{i+b} \text{, for each }i\in\{0,1,...,n-1\}.
\end{equation}
Moreover, the left branch of input continuous difference of the next round, denoted by $x^{r+2}$, is generated by the XOR operation of $y^{r}$, $x^{r}$ and $S^c(x^{r+1})$. According to the XOR rule in Proposition~\ref{prop:CF-XOR-rule}, $x^{r+2}_i$ can be easily computed by $x^{r+2}_i = y^{r}_i \cdot x^{r}_{i}\cdot x^{r+1}_{i+c}$. However, \textsf{Gurobi} does not support the cubic or higher-degree constraints. Hence, we need to introduce an auxiliary vectorial variable $t^r\in\mathbb{R}^n$ as a temporary to represent the XOR of $y^{r}$ and $x^{r+1}$, then describe them as
\begin{equation}\label{equ:model_XOR-CD}
		t^{r}_i = y^{r}_i \cdot x^{r}_{i}\text{ and }
		x^{r+2}_i = t^{r}_i \cdot x^{r+1}_{i+c}\text{, for each $i$ in $\{0,...,n-1\}$}.
\end{equation}
Note that the initial continuous difference of each input bit of the middle part are determined by the concrete output difference of the differential part. Assume that the left and right branches of output difference for the $R_d$-round differential part are denoted by $n$-bit variables $\alpha^{R_d+1}$ and $\alpha^{R_d}$, then the initial continuous difference of middle part, denoted by $x^1$ and $x^0$, are described as 
\begin{equation}\label{equ:model_initial_CD}
	 x^1_i + 2\alpha^{R_d+1}_i = 1 \text{ and }  x^0_i + 2\alpha^{R_d}_i = 1 \text{, for each $i\in\{0,...,n-1\}$}.
\end{equation}
This is because $x^1_i$ (resp. $x^0_i$) is 1 if $\alpha^{R_d+1}_i = 0$ (resp. $\alpha^{R_d}$), otherwise it is -1.
Also, the final correlation of middle part depends on the concrete initial mask of the linear part. Let $\lambda^0$ and $\lambda^1$ denote the left and right branches of the initial masks of the linear part, respectively. The correlation of the $R_m$-round middle part, denoted by $Cor_m$ where its value is between -1 and 1, can represented as
\[
Cor_m = (\prod_{i=0, \lambda^0_i=1}^{n-1}x^{R_m+1}_i) \cdot (\prod_{i=0, \lambda^1_i=1}^{n-1}x^{R_m}_i)
\]  
according to Proposition~\ref{prop:CF-XOR-rule}. It is too difficult to encode above conditional continuous products. However, taking the logarithm on the absolute value of $Cor_m$, we can transform the continuous product to the following quadratic expression
\[
\log_2|Cor_m| = \sum_{i=0}^{n-1}(\lambda^0_i\cdot\log_2|x^{R_m+1}_i| + \lambda^1_i\cdot\log_2|x^{R_m}_i|).
\] 
Therefore, modeling above expression needs to consider how to constrain the absolute and logarithmic functions with MILP or MIQCP. In~\cite{DBLP:conf/ctrsa/BelliniGGMP23}, Bellini \textit{et al.} presented a piece-wise linear function $g(x)$ (see Appendix~\ref{appendix:B}) to approximate $-\log_2|x|$ such that $g(x) \leq -\log_2|x|$. They implemented the modeling process with MiniZinc and solved the model also by \textsf{Gurobi}. Similarly, Zhou \textit{et al.}~\cite{journals/iet/ZhouWH24} adopted the same way to construct model for \textsf{Simon}-like ciphers.  

Fortunately, \textsf{Gurobi} supports the absolute and logarithmic functions interfaces for some programming languages like \textsf{Python}, \textsf{C/C++}, \textsf{Matlab}, etc. These integrated functions are called \textit{general expressions}\footnote{For more details, please refer to the \textsf{Gurobi} optimizer reference manual at \url{https://www.gurobi.com/documentation/current/refman/index.html}} in \textsf{Gurobi}. In this paper, we utilize the \textsf{Python} interfaces `$A$=\texttt{abs\_(}$B$\texttt{)}' and `\texttt{addGenConstrLogA(}$A, B, C$\texttt{)}' to straightforwardly achieve the descriptions: $A = |B|$ and $B=\log_{C}(A)$, where $A,B,C$ belong to $\mathbb{R}$.
To describe aforementioned relation, we here introduce an auxiliary variable $\mathtt{Cor}_m\in\mathbb{R}$ to represent $\log_2|Cor_m|$, and four vectorial variables $z^0$, $z^1$, $z^2$, $z^3$ to represent $|x^{R_m+1}|$, $|x^{R_m}|$, $\log_2|x^{R_m+1}|$, $\log_2|x^{R_m}|$, respectively. Thus the logarithm on the absolute value of the correlation in middle part can be constrained by the following general expressions
\begin{equation}\label{equ:model_middle_Cor}
	\begin{cases}
		z^0_i = \texttt{abs\_(}x^{R_m+1}_i\texttt{)}\\
		z^1_i = \texttt{abs\_(}x^{R_m}_i\texttt{)}\\
		\texttt{addGenConstrLogA(}z^0_i, z^2_i, 2\texttt{)}\\
		\texttt{addGenConstrLogA(}z^1_i, z^3_i, 2\texttt{)}\\
		\text{for each $i\in\{0,...,n-1\}$ }\\
		\texttt{Cor}_m = \sum_{j=0}^{n-1}(\lambda^0_j\cdot z_j^2+\lambda^1_j\cdot z^3_j).
	\end{cases}
\end{equation}
It is worth noting that if $z^0_i = 0$ (resp. $z^1_i=0$) then $z^2_i = \log_2 z^0_i$ (resp. $z^3_i = \log_2 z^1_i$) will be meaningless. However, in $\mathsf{Gurobi}$, the value of $z^2_i$ or $z^3_i$ will be signed to `$\mathtt{-inf}$', i.e., negative infinity. Thus it will take no effect on solving the model if our objective function is to minimize the value of $-\mathtt{Cor}_m$ (note $\mathtt{Cor}_m$ is negative).
 
In summary, combing Equation~\eqref{equ:model_AND-CD}-\eqref{equ:model_middle_Cor}, we can use an MIQCP model that contains about $(3n\cdot R_m + 6n + 1)$ constraints and $(3n\cdot R_m+6n+R_m)$ variables, to describe the $R_m$-round middle part. The process is depicted in Algorithm~\ref{alg:model-middle-part}.

Compared to the modeling way used in~\cite{DBLP:conf/ctrsa/BelliniGGMP23} and~\cite{journals/iet/ZhouWH24}, ours is more superior. On the one hand, our implementation is quite easy and intuitive. On the another hand, the use of approximate function $g(x)$ may lead to an inaccurate result. In particular, the approximate value will be less than the exact value. Because $\min\{g(x)\}\leq \min\{-\log_2|x|\}$ holds and the goal of model is to minimize the objective function. 

\paragraph{Modeling the linear part with precision.} K\"{o}lbl \textit{et al.}~\cite{DBLP:conf/crypto/KolblLT15} investigated the exact linear behavior of \textsf{Simon}-like round function, and presented an efficient computation for calculating the correlation as shown in Algorithm~\ref{alg:calculate-cor} of Appendix~\ref{appendix:algorithms}. Next we detail how to use MILP method to encode the computation of correlation for one round. For the $r$th round linear part, we denote $(\lambda^{r},\lambda^{r+1})$ and $(\lambda^{r+1},\lambda^{r+2})$ the input and output masks, respectively, where $\lambda^{r},\lambda^{r+1}$ and $\lambda^{r+2}$ belong to $\mathbb{F}_2^n$. Denote $\lambda^{in,r}$ the input mask generated by AND operation, then the following linear inequalities can describe line 2-5 of Algorithm~\ref{alg:calculate-cor}. 
\begin{equation}\label{equ:model-lambda_in}
	\begin{cases}
		\lambda^{r}_i + \lambda^{r+1}_{i-c} + \lambda^{r+2}_i - \lambda^{in,r}_i \geq 0\\
		\lambda^{r}_i + \lambda^{r+1}_{i-c} - \lambda^{r+2}_i + \lambda^{in,r}_i \geq 0\\
		\lambda^{r}_i - \lambda^{r+1}_{i-c} + \lambda^{r+2}_i + \lambda^{in,r}_i \geq 0\\
		-\lambda^{r}_i + \lambda^{r+1}_{i-c} + \lambda^{r+2}_i + \lambda^{in,r}_i \geq 0\\
		\lambda^{r}_i + \lambda^{r+1}_{i-c} + \lambda^{r+2}_i - \lambda^{in,r}_i \leq 2\\
		\lambda^{r}_i + \lambda^{r+1}_{i-c} - \lambda^{r+2}_i + \lambda^{in,r}_i \leq 2\\
		\lambda^{r}_i - \lambda^{r+1}_{i-c} + \lambda^{r+2}_i + \lambda^{in,r}_i \leq 2\\
		-\lambda^{r}_i + \lambda^{r+1}_{i-c} + \lambda^{r+2}_i + \lambda^{in,r}_i \leq 2\\
		\lambda^{r+1}_{i-a} + \lambda^{r+1}_{i-b} - \lambda^{in,r}_i \geq 0\\
		\text{for each $i\in\{0,1,...,n-1\}$}.
	\end{cases} 	
\end{equation}
Note that we expect to find the high correlation, so we do not need to consider the case of $\lambda^{r+1}=1^n$. In other words, we need to constrain the weight of $\lambda^{r+1}$ less than $n
$ as 
\begin{equation}\label{equ:model-lambda_out}
	\sum_{i=0}^{n-1}\lambda^{r+1}_i\leq n-1.
\end{equation}
For the line 19-23 of Algorithm~\ref{alg:calculate-cor}, $\mathtt{tmp}$ is updated by $\bigwedge_{j=0}^{j=k}S^{j\cdot(b-a)}(\lambda^{r+1})$ at the $k$th loop and the loop will be broken when $\mathtt{tmp}=0^n$. Note that $\lambda^{r+1}\neq 1^n$, thus $\mathtt{tmp}$ must equal to $0^n$ after at most $n-1$ loops. We introduce $n$ binary variable lists $\mathtt{tmp}^{0,r},...,\mathtt{tmp}^{n-1,r}$ where $\mathtt{tmp}^{0,r}=\lambda^{r+1}$. Here, we utilize the general expression `$A=\texttt{and\_(}B,C\texttt{)}$’ offered by \textsf{Gurobi} to directly describe the relation $A = B\wedge C$, where $A,B,C$ are binary variables. Thus, the following general expression
\begin{equation}\label{equ:model-tmp}
	\mathtt{tmp}^{j+1,r}_i = \texttt{and\_(}\mathtt{tmp}^{j,r}_i, \lambda^{r+1}_{i+j\cdot(b-a)}\texttt{)}\text{, for each $j\in\{0,...,n-2\}$, $i\in\{0,...,n-1\}$}
\end{equation}
can be used to describe the update on the value of $\mathtt{tmp}$. Moreover, $\mathtt{abits}$ is equal to the continuous XOR with current $\mathtt{tmp}$, i.e., $\mathtt{abits}^r$ = $\bigoplus_{j=0}^{n-1}\mathtt{tmp}^{j,r}$ which can be constrained by
\begin{equation}\label{equ:model-abits}
	\sum_{j=0}^{n-1}\mathtt{tmp}^{j,r}_i + \texttt{abits}^r_i = 2N^r_i \text{, for each $i\in\{0,...,n-1\}$}
\end{equation}
where $N^r\in\mathbb{Z}^n$. Similarly, we can constrain $\mathtt{sbits}$ and $\mathtt{pbits}$ by
\begin{equation}\label{equ:model-pbits}
	\begin{cases}
		- \mathtt{abits}^r_{i-a} - \mathtt{sbits}^{0,r}_i \geq -1\\
		- \lambda^{r+1}_{i-b} - \mathtt{sbits}^{0,r}_i \geq -1\\
		- \lambda^{r+1}_{i-a} + \lambda^{r+1}_{i-b} + \mathtt{abits}^r_{i-a} + \mathtt{sbits}^{0,r}_i \geq 0\\
		\lambda^{r+1}_{i-a} - \mathtt{sbits}^{0,r}_i \geq 0\\
		\mathtt{pbits}^{0,r}_{i-2a+2b} = \texttt{and\_(sbits}^{0,r}_{i}, \lambda^{in,r}_{i}\texttt{)}\\
		\quad
		\begin{cases}
			\mathtt{sbits}^{j+1,r}_{i-2a+2b} = \texttt{and\_(sbits}^{j,r}_{i}, \lambda^{r+1}_{i-a}\texttt{)}\\
			\lambda^{in,r}_{i} - \mathtt{pbits}^{j,r}_{i} + \mathtt{pbits}^{j+1,r}_{i-2a+2b} \geq 0\\
			\lambda^{in,r}_{i} + \mathtt{pbits}^{j,r}_{i} - \mathtt{pbits}^{j+1,r}_{i-2a+2b} \geq 0\\
			-\mathtt{sbits}^{j,r}_{i}- \lambda^{in,r}_{i} + \mathtt{pbits}^{j,r}_{i} + \mathtt{pbits}^{j+1,r}_{i-2a+2b} \geq -1\\
			-\mathtt{sbits}^{j,r}_{i}- \lambda^{in,r}_{i} - \mathtt{pbits}^{j,r}_{i} - \mathtt{pbits}^{j+1,r}_{i-2a+2b} \geq -3\\
			\mathtt{sbits}^{j,r}_{i} - \mathtt{pbits}^{j,r}_{i} + \mathtt{pbits}^{j+1,r}_{i-2a+2b} \geq 0\\
			\mathtt{sbits}^{j,r}_{i} + \mathtt{pbits}^{j,r}_{i} - \mathtt{pbits}^{j+1,r}_{i-2a+2b} \geq 0\\
			\text{for each $j\in\{0,...,n-2\}$}
		\end{cases}\\
		\mathtt{pbits}^{n-1,r}_i = 0\\
		\text{for each $i\in\{0,...,n-1\}$}.
	\end{cases}
\end{equation}
Finally, the logarithm on the absolute value of correlation, denoted by $\mathtt{Cor}^r\in\mathbb{Z}$, is equal to the weight of $\mathtt{abits}^r$ and described as
\begin{equation}\label{equ:model-linear-cor}
	\mathtt{Cor}^r = \sum_{i=0}^{n-1}\texttt{abits}^r_i.
\end{equation}
Therefore, there are total $(8n^2+8n+2)$ constraints and $(3n^2+4n+1)$ variables by combing Equation~\eqref{equ:model-lambda_in}-\eqref{equ:model-linear-cor} to describe the correlation for one round linear mask propagation. The modeling process of $R_l$-round linear part is depicted in Algorithm~\ref{alg:model-linear-part}.

In~\cite{journals/iet/ZhouWH24}, Zhou \textit{et al.} regarded the AND operations as independent S-boxes and described its linear approximation table (LAT) to constrain the relation between the input-output masks and the correlation. Their model is very simple but quite imprecise. For an instance of 4-round linear approximation of \textsf{Simon32}, we set the output mask as $(\mathtt{0x40,0x10})$, and use our model and Zhou \textit{et al.}'s model to compute the number of possible input masks, respectively. The result is listed in the below table. Apparently, the imprecise model actually misses a part of valid linear input masks, which will influence the estimated correlation of DL distinguisher when taking the clustering effect into consideration.
\begin{table}[!h]
	\centering
	\caption{The number of possible input masks that propagate to $(\mathtt{0x40,0x10})$ through 4-round \textsf{Simon32}. $q$: linear approximation correlation. Model 1: Zhou \textit{et al.} model. Model 2: our new model.}
	\begin{tabular}{ccccccccc}
		\toprule
		$\log_2q$ & $-1$ & $-2$ & $-3$ & $-4$ & $-5$ & $-6$ & $-7$ & $-8$ \\ 
		Model 1 & 0 & 0 & 4 & 32 & 52 & 212 & 848 & 3192 \\ 
		Model 2 & 0 & 0 & 4 & 32 & 88 & 1328 & $>5500$ & $>8000$ \\ 
		\bottomrule
	\end{tabular}
\end{table}
\vspace{-0.5cm}
\paragraph{Merging the three parts.} Assuming the $R$-round target cipher is divided to $R_d$-round differential part, $R_m$-round middle part and $R_l$-round linear part where $R=R_d+R_m+R_l$. Then, the sub-model corresponds to each part can be constructed by performing the procedures \texttt{ModelDiffPart}, \texttt{ModelMiddlePart} and \texttt{ModelLinearPart} (see in Algorithm~\ref{alg:model-diff-part},~\ref{alg:model-middle-part} and~\ref{alg:model-linear-part} of Appendix~\ref{appendix:algorithms}). Finally, we merge the three sub-models and set an appropriate objective function to generate an entire model for finding the best DL trail. Algorithm~\ref{alg:model-entire} in Appendix~\ref{appendix:algorithms} briefly shows how to find an $R$-round DL trail with a given round configuration using MILP/MIQCP method. 

\subsection{Finding DL Trails with Two Heuristic Strategies}
Note that the precondition that Algorithm~\ref{alg:model-entire} can work is the round configuration $(R_d,R_m,R_l)$ is fixed. In other words, it is inevitable to traverse all the possible configurations if we want to find the best $R$-round DL trail. Nevertheless, it is not practical because an entire model contains about $(8n^2\cdot R_l + 3n\cdot R+19n\cdot R_d+5n\cdot R_l+6n)$ constraints and $(3n^2\cdot R_l+3n\cdot R+3n\cdot R_d+n\cdot R_l+10n)$ variables. Also, the types of these constraints and variables are sundry. Thus when the round and block size increase, it will be quite difficult to solve one model. For instance, with regard to the 15-round $\textsf{Simon48}$ with the round configuration $(5,5,5)$, we send the entire 15-round model into $\mathsf{Gurobi}$ to solve. However, after 24 hours, the solving process is still going on. At present, the current candidate solution is $19.49$ and the optimal solution is lower bounded by $7.93$, which indicates that the solving gap is $59.3\%$. It is difficult to evaluate how long it will take for \textsf{Gurobi} to return the final solution of this model, thus we interrupt the procedure in advance and get a DL trail with correlation $2^{-19.49}$. 

After performing plenty of experiments on \textsf{Simon32} with different configurations, we find that in most cases, these obtained DL trails consist of either the best differential characteristics or linear characteristics. From this observation, we empirically deduce that the best DL trail probably contains the best differential or linear characteristic. This inspires us to speed up the automatic search in a heuristic way, i.e., we can search for the DL trail from the best differential or linear characteristic. For this, we adopt the divide-and-conquer idea and present the differential-first and linear-first strategies as below.
\begin{itemize}
	\item[-] The differential-first strategy (DFS). This strategy splits the entire process into three steps. First, it requires to construct an independent model for the $R_d$-round differential part, and solve it to obtain the best differential characteristic. Under this, the next step is to transform the obtained output difference to input continuous difference of the $R_m$-round middle part and call Algorithm~\ref{alg:calculate-continuous-diff} to calculate the output continuous difference of middle part. At last, we only need to construct an $R_l$-round model that describes not only the correlation derived by the $R_l$-round linear approximation but also the one derived by selecting the mask on the $R_m$-round output continuous difference. Thus, solving the $R_l$-round model can obtain a DL trail. Algorithm~\ref{alg:differential-first-strategy} illustrates the process of finding DL trails with this strategy.
	\item[-] The linear-first strategy (LFS). This strategy consists of two steps. First, we construct an independent model for $R_l$-round linear part, and solve it to obtain the best linear characteristic. Under the condition that best $R_l$-round linear correlation and trail are known, the next step is to construct models for $R_d$-round differential and $R_m$-round middle parts and merge them to an entire model. Similarly, we need to append additional constraints and set an appropriate objective function for this entire model. As a result, a DL trail can be derived by solving the model. We clarify the process in Algorithm~\ref{alg:linear-first-strategy}. 
\end{itemize}

The above two strategies aim to divide the entire but unsolvable model into two small-scale sub-models, which can be solved rapidly. Also, DFS is superior to LFS from the efficiency perspective. Because under the same configuration, DFS has the smaller scale models and the constraints are purely linear. We apply our strategies to the aforementioned 15-round \textsf{Simon48}, it only takes about 40 seconds to get a better DL trail with correlation $2^{-18.66}$, which improves the efficiency significantly compared to solving the entire model.  We stress that the two strategies are heuristic, thus the corresponding correlations both are actually lower than the one obtained by the entire model in theory. Also, the correlations derived by different strategies may be distinct. Therefore, regarding a round configuration, it is necessary to try both of the two strategies to obtain the better result. 

\subsection{Deriving DL Distinguishers by Transforming Technique}\label{Sect:3.3}
As mentioned in Section~\ref{Sect:2.2}, the previous automatic method~\cite{DBLP:conf/ctrsa/BelliniGGMP23,journals/iet/ZhouWH24} only aims to find a single DL trail with a high correlation that corresponds to Equation~\eqref{equ:DL-trail}. However, the estimated correlation of a single DL trail may be far (higher or lower) from that of a real distinguisher. Therefore, it is of great significance to explore the valuable DL trails, which share the common difference-mask pair and contribute a lot to the final correlation, to derive a DL distinguisher. This is similar to exploiting the clustering effect on differential characteristics to improve the probability in the classical differential cryptanalysis. 

Here, we propose the \textit{transforming technique}, which means transforming a given DL trail to a real DL distinguisher, to explore the potential DL distinguishers. According to Equation~\eqref{equ:DLCT-formula}, the transforming technique seems trivial but indeed effective with the help of automatic tool. Assume that there is a given DL trail that its information including its differential-mask $(\Delta_I,\lambda_O)$, the differential probability $p$, the linear approximation correlation $q$ and the round configuration $(R_d,R_m,R_l)$ are known, then the transforming technique to find a distinguisher based on this given trail is detailed as below:
\begin{enumerate}
	\item For the $R_d$-round differential part $E_d$, we enumerate the possible differential trails that start from the input difference $\Delta_I$. This can be also implemented efficiently by automatic tools like MILP described in Algorithm~\ref{alg:model-diff-part} or SMT/SAT used in~\cite{DBLP:conf/crypto/KolblLT15}. Assume that there exists a differential trail $\Delta_I\overset{E_d}{\rightarrow}\Delta_*$ with the probability of $p_*$, then we record the pair $(\Delta_*,p_*)$ and add it into the list $\mathbb{L}_d$.
	\item Similarly, for the $R_l$-round linear part $E_l$, we enumerate the possible linear trails that end up with $\lambda_O$ by the automatic method. Assume that there exists a linear trail $\lambda_*\overset{E_l}{\rightarrow}\lambda_O$ with the correlation of $q_*$, then we record the pair $(\lambda_*,q_*)$ and add it into the list $\mathbb{L}_l$.
	\item Select the $i$th pair $(\Delta_i,p_i)$ from $\mathbb{L}_d$ and the $j$th pair $(\lambda_j,q_j)$ from $\mathbb{L}_l$, a correlation $r_{i,j}$ for the $R_m$-round middle part $E_m$ can be computed by Algorithm~\ref{alg:calculate-continuous-diff} and the correlation of this single DL trail $(\Delta_I,\Delta_i,\lambda_j,\lambda_O)$ for $E$ can be calculated as $p_i\cdot r_{i,j}\cdot q^2_j$. 
	\item Traverse all the pairs in $\mathbb{L}_d$ and $\mathbb{L}_l$, and sum over the computed correlations of all DL trails by repeating the step 3, the correlation of the real DL distinguisher $\Delta_I\overset{E}{\rightarrow}\lambda_O$ is estimated as $\sum_{i}\sum_j p_i\cdot r_{i,j}\cdot q^2_j$.
\end{enumerate}
Note that when the differential probability or linear approximation correlation is very small, the corresponding DL trail will not be significantly contributive to the correlation of DL distinguisher. Hence, we can set an appropriate lower bound on the probability (resp. correlation) in step 1 (resp. step 2) when searching for the differential (resp. linear) trails. This strategy can not only improve the efficiency of the whole process but also not lose the precision on the correlation of the transformed DL distinguisher. For example, a 13-round DL approximation (it is actually a DL trail) with correlation $2^{-16.63}$ for \textsf{Simon32} was presented in~\cite{journals/iet/ZhouWH24}. In particular, the round configuration is $(5,5,3)$, the 5-round differential probability is $2^{-8}$ and the 3-round linear approximation correlation is $2^{-4}$. By setting the lower bounds on probability and correlation as $2^{-16}$ and $2^{-8}$ respectively, it took 33 minutes to enumerate the DL trails and obtain a real DL distinguisher with the correlation of $2^{-13.94}$. Nevertheless, in the same experimental platform, when decreasing the lower bounds on probability and correlation to $2^{-18}$ and $2^{-9}$ respectively, it took 110 minutes but only had a very few improvement on the final correlation from $2^{-13.94}$ to $2^{-13.92}$. We illustrate the procedure how to implement transforming technique with the mentioned strategy using MILP method in Algorithm~\ref{alg:transforming-technique}.

 \section{Applications to \textsf{Simon} and \textsf{Simeck}}\label{Sect:4}
 In this section, we will apply our improved method to searching for better DL trails and further exploring DL distinguishers for \textsf{Simon} and \textsf{Simeck}. Also, we verify some DL distinguishers for the small-block versions by experiments. All the procedures and experiments in this paper are performed on a PC with Windows 11 system, Intel Core i7-8700 CPU, 24 GB RAM and $\mathsf{Gurobi}$ optimizer v10.0.2.
 
  \subsection{Finding Differential-Linear Trails}\label{Sect:4.1}
 By our precise model and the search strategies, we find the better DL trails for the all versions of \textsf{Simon} and \textsf{Simeck}. In Table~\ref{tab:DL-trails}, we list the brief information of our new DL trails and the comparison with~\cite{journals/iet/ZhouWH24}. 
 \begin{table}[!h]
 	\centering
 	\caption{The DL trails for the all versions of $\mathsf{Simon}$ and \textsf{Simeck}. \#R: the number of round. Config: the round configuration. \textsf{Cor}: the correlation. The new DL trails found in our work are bolder, others are previously presented by~\cite{journals/iet/ZhouWH24}. The specification on these DL trails can be found in Appendix~\ref{appendix:DL-trails}.}{\label{tab:DL-trails}}
 	\subtable{
 		\renewcommand\tabcolsep{6.5pt}
 		\begin{tabular}{cccc}
 			\toprule
 			Cipher&\#R&Config&\textsf{Cor}\\
 			\midrule
 			\multirow{4}{*}{\textsf{Simon32}}&13&$(5,5,3)$&$2^{-16.63}$\\
 			&\textbf{13}&$\bm{(5,5,3)}$&$\bm{2^{-14.73}}$\\
 			&14&$(5,5,4)$&$2^{-18.63}$\\
 			&\textbf{14}&$\bm{(5,6,3)}$&$\bm{2^{-17.49}}$\\\midrule
 			\multirow{7}{*}{\textsf{Simon48}}&15&$(7,4,4)$&$2^{-20.19}$\\
 			&\textbf{15}&$\bm{(5,5,5)}$&$\bm{2^{-18.66}}$\\
 			&16&$(7,5,4)$&$2^{-22.66}$\\
 			&\textbf{16}&$\bm{(7,5,4)}$&$\bm{2^{-21.30}}$\\
 			&\textbf{16}&$\bm{(5,6,5)}$&$\bm{2^{-21.01}}$\\
 			&17&$(7,6,4)$&$2^{-24.66}$\\
 			&\textbf{17}&$\bm{(6,6,5)}$&$\bm{2^{-24.08}}$\\\midrule
 			\multirow{3}{*}{\textsf{Simon64}}&20&$(7,7,6)$&$2^{-34.58}$\\
 			&\textbf{20}&$\bm{(7,7,6)}$&$\bm{2^{-33.06}}$\\
 			&\textbf{20}&$\bm{(8,6,6)}$&$\bm{2^{-32.46}}$\\\midrule
 			\multirow{4}{*}{\textsf{Simon96}}&25&$(10,6,9)$&$2^{-46.66}$\\
 			&\textbf{25}&$\bm{(9,8,8)}$&$\bm{2^{-44.41}}$\\
 			&26&$(11,6,9)$&$2^{-50.66}$\\
 			&\textbf{26}&$\bm{(9,9,8)}$&$\bm{2^{-50.16}}$\\
 			\bottomrule
 		\end{tabular}
 	}
 	\subtable{
 		\renewcommand\tabcolsep{6.2pt}
 		\begin{tabular}{cccc}
 			\toprule
 			Cipher&\#R&Config&\textsf{Cor}\\
 			\midrule
 			
 			\multirow{4}{*}{\textsf{Simon128}}&31&$(10,9,12)$&$2^{-62.70}$\\
 			&\textbf{31}&$\bm{(13,11,7)}$&$\bm{2^{-61.35}}$\\
 			&32&$(11,9,12)$&$2^{-66.70}$\\
 			&\textbf{32}&$\bm{(13,10,9)}$&$\bm{2^{-65.61}}$\\\midrule
 			\multirow{2}{*}{\textsf{Simeck32}}&14&$(5,5,4)$&$2^{-16.63}$\\
 			&\textbf{14}&$\bm{(5,6,3)}$&$\bm{2^{-15.99}}$\\\midrule
 			\multirow{5}{*}{\textsf{Simeck48}}&17&$(6,6,5)$&$2^{-22.37}$\\
 			&\textbf{17}&$\bm{(6,6,5)}$&$\bm{2^{-22.07}}$\\
 			&\textbf{17}&$\bm{(5,7,5)}$&$\bm{2^{-21.44}}$\\
 			&18&$(6,6,6)$&$2^{-24.75}$\\
 			&\textbf{18}&$\bm{(8,7,3)}$&$\bm{2^{-24.06}}$\\\midrule
 			\multirow{8}{*}{\textsf{Simeck64}}&22&$(7,7,8)$&$2^{-32.44}$\\
 			&\textbf{22}&$\bm{(4,9,9)}$&$\bm{2^{-29.04}}$\\
 			&23&$(7,7,9)$&$2^{-36.13}$\\
 			&\textbf{23}&$\bm{(4,10,9)}$&$\bm{2^{-32.44}}$\\
 			&24&$(7,7,10)$&$2^{-38.13}$\\
 			&\textbf{24}&$\bm{(4,9,11)}$&$\bm{2^{-35.04}}$\\
 			&25&$(7,7,11)$&$2^{-41.04}$\\
 			&\textbf{25}&$\bm{(4,10,11)}$&$\bm{2^{-39.04}}$\\
 			\bottomrule
 		\end{tabular}
 		
 	}
 \end{table} 
 
The above tables show that most of our DL trails have the different round configurations but there exist a few ones that have the same round configurations as in~\cite{journals/iet/ZhouWH24} including the 13-round \textsf{Simon32}, 16-round \textsf{Simon48}, 20-round \textsf{Simon64} and 17-round \textsf{Simeck48}. Thereby, under the same configurations, we can also explore the DL trail with a higher correlation. This improvement mainly benefits from the DFS and DLS search methods. For example, under the configuration $(5,5,3)$ of 13-round \textsf{Simon32}, we find a better DL trail with the correlation of $2^{-14.74}$, which improves the previous best one $2^{-16.73}$ by a factor of 4.

\subsection{Finding Differential-Linear Distinguishers}\label{Sect:4.2}
For finding DL distinguishers with transforming technique, we prepare a lot of DL trails including the best ones listed in Table~\ref{tab:DL-trails} as well as some newly found trails with very low correlations. As a result, we not only improve the correlations of previous DL distinguishers but also find some better ones. Here we only list the DL distinguisher with the best correlation in Table~\ref{tab:best-DL-distinguisher}. Among these, the DL distinguishers of \textsf{Simon} have the best correlations so far. Especially, we extend the previous best theoretical DL distinguishers for \textsf{Simon32/48/96} by one round and for \textsf{Simon64} by two round. 
\begin{table}[!h]
	\centering
	\caption{The best distinguishers of \textsf{Simon} and \textsf{Simeck}. \#R: the number of round. \textsf{Cor}: the correlation of distinguisher. The details of these DL trails can be found in Appendix~\ref{appendix:DL-trails}.}{\label{tab:best-DL-distinguisher}}
	\renewcommand\tabcolsep{4.0pt}
	\begin{threeparttable}
			\begin{tabular}{c|c|c|c|c}
				\toprule
				Cipher & \#R & Input difference & Output mask & \textsf{Cor} \\\midrule
				\multirow{3}{*}{\textsf{Simon32}}
				& 13 & $\mathtt{(0x100,0x440)}$ & $(\mathtt{0x800,0x2200})$ & $2^{-11.99}$ \\
				& 14 & $\mathtt{(0x100,0x645)}$ & $(\mathtt{0x8000,0x2002})$ & $2^{-15.36}$ \\
				& 15 & $\mathtt{(0x8,0x22)}$ & $(\mathtt{0x4,0x111})$ & $2^{-18.29}$ \\
				\midrule
				\multirow{4}{*}{\textsf{Simon48}}
				& 15 & $(\mathtt{0x800},\mathtt{0x2220})$ & $(\mathtt{0x4},\mathtt{0x11})$ & $2^{-15.70}$ \\
				& 16 & $(\mathtt{0x400},\mathtt{0x1110})$ & $\mathtt{(0x8, 0x22)}$ & $2^{-18.90}$ \\
				& 17 & $(\mathtt{0x80,0x222})$ & $(\mathtt{0x40000,0x110001})$ & $2^{-21.88}$ \\
				& 18 & $(\mathtt{0x80, 0x222})$ & $(\mathtt{0x8, 0x222})$ & $2^{-24.55}$ \\
				\midrule
				\multirow{2}{*}{\textsf{Simon64}}
				& 20 & $(\mathtt{0x800},\mathtt{0x83220})$ & $(\mathtt{0x8},\mathtt{0x222})$ & $2^{-28.67}$ \\
				& 21 & $(\mathtt{0x800},\mathtt{0x2220})$ & $(\mathtt{0x20},\mathtt{0x888})$ & $2^{-31.99}$\\
				\midrule
				\multirow{2}{*}{\textsf{Simon96}}
				& 25 & $(\mathtt{0x800,0x2220})$ & $(\mathtt{0x20,0x888})$& $2^{-40.85}$ \\
				& $26^{\dagger}$ & $(\mathtt{0x101000,0x440400})$ & $(\mathtt{0x20,0x888})$ & $2^{-44.03}$ \\
				\midrule
				\multirow{2}{*}{\textsf{Simon128}}
				& 31 & $(\mathtt{0x8000000,0x22202000})$ & $(\mathtt{0x200,0x8880)}$ & $2^{-55.66}$ \\
				& 32 & $(\mathtt{0x4040000,0x11010000})$ & $(\mathtt{0x8,0x20222})$ & $2^{-59.62}$ \\
				\midrule
				\textsf{Simeck32} & 14 & $(\mathtt{0x4,0x800a})$ & $(\mathtt{0x2000,0x5000})$ & $2^{-14.73}$ \\
				\midrule
				\multirow{3}{*}{\textsf{Simeck48}}
				& 17 & ($\mathtt{0x8000,0x15000}$) & $(\mathtt{0x200,0x500})$ & $2^{-17.79}$ \\
				& 18 & ($\mathtt{0x8000,0x14000}$) & $(\mathtt{0x400,0x2a00})$ & $2^{-20.41}$ \\
				& 19 & ($\mathtt{0x8000,0x15000}$) & $(\mathtt{0x80,0x540})$ & $2^{-23.22}$ \\
				\midrule
				\multirow{4}{*}{\textsf{Simeck64}}
				& 22 & $(\mathtt{0x20000,0x54000})$ & $(\mathtt{0x400,0x2a00})$ & $2^{-25.98}$ \\
				& 23 & $(\mathtt{0x100,0x280})$ & $(\mathtt{0x0,0x44})$ & $2^{-30.92}$ \\
				& 24 & $(\mathtt{0x0,0x880})$ & $(\mathtt{0x8,0x54})$ & $2^{-31.88}$ \\
				& 25 & $(\mathtt{0x10,0x28})$ & $(\mathtt{0x4,0x2})$ & $2^{-35.53}$ \\
				\bottomrule
			\end{tabular}
		\begin{tablenotes}
			\footnotesize
			\item[$\dagger$] The trail that used to derive this distinguisher comes from Zhou \textit{et al.}'s work~\cite{journals/iet/ZhouWH24}, others are newly found in our work.
		\end{tablenotes}
	
	\end{threeparttable}
\end{table}
  
 In addition, in order to intuitively show the effectiveness of transforming technique, we list the comparison on correlations of DL trails and the corresponding DL distinguishers in Table~\ref{tab:DL-distinguisher-simon} and~\ref{tab:DL-distinguisher-simeck}. We use the improvement percentage, calculated by (\textsf{Cor$_D$$-$Cor$_T$})/\textsf{Cor$_T$}, to describe the gap on correlations of a DL trail and its derived DL distinguisher. From the two tables, we can see that there actually exists a noticeable gap on the correlations of a DL trail and its derived DL distinguisher. Also, the correlation of a distinguisher is always higher than that of a trail. Namely, the transforming technique actually makes sense to improve the correlation. Interestingly, some DL trails have the very low correlations but can derive the high-correlation DL distinguishers. For an instance of the 14-round \textsf{Simon32}, there are three DL trails with correlations of $2^{-18.63}$, $2^{-17.49}$ and $2^{-20}$, respectively. However, the corresponding DL distinguishers derived by the three trails have the correlations of $2^{-15.63}$, $2^{-16.75}$ and $2^{-15.36}$, where the highest-correlation trail derives the lowest-correlation distinguisher but the lowest-correlation trail derives the highest-correlation distinguisher. To exhibit these phenomena more directly, we highlight the best correlations respectively for trails and distinguishers with the bolder blue and red in Table~\ref{tab:DL-distinguisher-simon} and~\ref{tab:DL-distinguisher-simeck}.
 
 \begin{table}[!h]
 	\centering
 	\caption{The comparison on correlations of DL trails and DL distinguishers of \textsf{Simon}. \textsf{Cor$_T$}: the correlation of DL trail. \textsf{Cor$_D$}: the correlation of DL distinguisher. The details of these DL trails can be found in Appendix~\ref{appendix:DL-trails}.}{\label{tab:DL-distinguisher-simon}}
 	\begin{tabular}{c|c|c|c|c|c|c}
 		\toprule
 		Version&Round&Config&\textsf{Cor$_T$}&\textsf{Cor$_D$}&Gap&Source\\
 		\midrule
 		\multirow{8}{*}{\textsf{Simon32}}
 		&11&$(5,2,4)$&$2^{-14.00}$&$2^{-8.03}$&$6168\%$&This work\\\cmidrule{2-7}
 		&\multirow{3}{*}{13}
 		&$(5,5,3)$&$2^{-16.63}$&$2^{-13.92}$&$350\%$&~\cite{journals/iet/ZhouWH24}\\
 		&&$(5,5,3)$&\textcolor{blue}{$\bm{2^{-14.73}}$}&$2^{-12.61}$&$334\%$&This work\\
 		&&$(5,3,5)$&$2^{-16.83}$&\textcolor{red}{$\bm{2^{-11.99}}$}&$2764\%$&This work\\\cmidrule{2-7}
 		&\multirow{3}{*}{14}
 		&$(5,5,4)$&$2^{-18.63}$&$2^{-15.63}$&$700\%$&\cite{journals/iet/ZhouWH24}\\
 		&&$(5,6,3)$&\textcolor{blue}{$\bm{2^{-17.49}}$}&$2^{-16.75}$&$67\%$&This work\\
 		&&$(7,3,4)$&$2^{-20.00}$&\textcolor{red}{$\bm{2^{-15.36}}$}&$2393\%$&This work\\\cmidrule{2-7}
 		&15&$(5,5,5)$&$2^{-20.72}$&$2^{-18.29}$&$438\%$&This work\\\midrule
 		\multirow{10}{*}{\textsf{Simon48}}
 		&14&$(5,4,5)$&$2^{-16.58}$&$2^{-13.33}$&$851\%$&This work\\\cmidrule{2-7}
 		&\multirow{3}{*}{15}
 		&$(7,4,4)$&$2^{-20.19}$&$2^{-16.83}$&$926\%$&\cite{journals/iet/ZhouWH24}\\
 		&&$(5,5,5)$&\textcolor{blue}{$\bm{2^{-18.66}}$}&$2^{-16.35}$&$395\%$&This work\\
 		&&$(7,3,5)$&$2^{-22.00}$&\textcolor{red}{$\bm{2^{-15.70}}$}&$7779\%$&This work\\
 		\cmidrule{2-7}
 		&\multirow{2}{*}{16}
 		&$(7,5,4)$&$2^{-22.66}$&$2^{-19.91}$&$572\%$&\cite{journals/iet/ZhouWH24}\\
 		&&$(7,5,4)$&\textcolor{blue}{$\bm{2^{-21.30}}$}&\textcolor{red}{$\bm{2^{-18.90}}$}&$427\%$&This work\\
 		
 		\cmidrule{2-7}
 		&\multirow{3}{*}{17}
 		&$(7,6,4)$&$2^{-24.66}$&$2^{-21.91}$&$572\%$&\cite{journals/iet/ZhouWH24}\\
 		&&$(6,6,5)$&\textcolor{blue}{$\bm{2^{-24.08}}$}&$2^{-22.52}$&$194\%$&This work\\
 		&&$(7,4,6)$&$2^{-26.00}$&\textcolor{red}{$\bm{2^{-21.88}}$}&$1638\%$&This work\\
 		\cmidrule{2-7}
 		&18&$(7,4,7)$&$2^{-28.58}$&$2^{-24.55}$&$1533\%$&This work\\
 		\midrule
 		\multirow{7}{*}{\textsf{Simon64}}
 		&16&$(5,6,5)$&$2^{-18.59}$&$2^{-16.26}$&$402\%$&This work\\\cmidrule{2-7}
 		&\multirow{4}{*}{20}
 		&$(7,7,6)$&$2^{-34.58}$&$2^{-31.80}$&$586\%$&\cite{journals/iet/ZhouWH24}\\
 		&&$(7,7,6)$&$2^{-33.05}$&$2^{-31.02}$&$308\%$&This work\\
 		&&$(8,6,6)$&\textcolor{blue}{$\bm{2^{-32.46}}$}&$2^{-29.52}$&$667\%$&This work\\
 		&&$(9,6,5)$&$2^{-33.16}$&\textcolor{red}{$\bm{2^{-28.67}}$}&$2147\%$&This work\\\cmidrule{2-7}
 		&21&$(8,5,8)$&$2^{-36.91}$&$2^{-31.99}$&2927\%&This work\\
 		\midrule
 		\multirow{5}{*}{\textsf{Simon96}}
 		&23&$(9,5,9)$&$2^{-40.00}$&$2^{-31.52}$&$35605\%$&This work\\\cmidrule{2-7}
 		&\multirow{2}{*}{25}
 		&$(10,6,9)$&$2^{-46.66}$&$2^{-41.59}$&$3259\%$&\cite{journals/iet/ZhouWH24}\\
 		&&$(9,8,8)$&\textcolor{blue}{$\bm{2^{-44.41}}$}&\textcolor{red}{$\bm{2^{-40.85}}$}&$1079\%$&This work\\\cmidrule{2-7}
 		&\multirow{2}{*}{26}
 		&$(11,6,9)$&$2^{-50.66}$&\textcolor{red}{$\bm{2^{-44.03}}$}&$9805\%$&\cite{journals/iet/ZhouWH24}\\
 		&&$(9,9,8)$&\textcolor{blue}{$\bm{2^{-50.16}}$}&$2^{-47.35}$&$601\%$&This work\\
 		\midrule
 		\multirow{5}{*}{\textsf{Simon128}}
 		&\multirow{3}{*}{31}
 		&$(10,9,12)$&$2^{-62.70}$&$2^{-56.86}$&$5628\%$&\cite{journals/iet/ZhouWH24}\\
 		&&$(13,11,7)$&\textcolor{blue}{$\bm{2^{-61.35}}$}&$2^{-57.91}$&$985\%$&This work\\
 		&&$(13,9,9)$&$2^{-61.73}$&\textcolor{red}{$\bm{2^{-55.66}}$}&6618\%&This work\\
 		\cmidrule{2-7}
 		&\multirow{2}{*}{32}
 		&$(11,9,12)$&$2^{-66.70}$&$2^{-60.03}$&$10082\%$&\cite{journals/iet/ZhouWH24}\\
 		&&$(13,10,9)$&\textcolor{blue}{$\bm{2^{-65.61}}$}&$2^{-61.61}$&$1500\%$&This work\\
 		&&$(11,8,13)$&$2^{-68.22}$&\textcolor{red}{$\bm{2^{-59.62}}$}&$38702\%$&This work\\
 		\bottomrule
 	\end{tabular}
 \end{table}

 \begin{table}[!h]
 	\centering
 	\caption{The comparison on correlations of DL trails and DL distinguishers for the all versions of \textsf{Simeck}. \textsf{Cor$_T$}: the correlation of DL trail. \textsf{Cor$_D$}: the correlation of DL distinguisher. The details of these DL trails can be found in Appendix~\ref{appendix:DL-trails}.}{\label{tab:DL-distinguisher-simeck}}
 	\begin{tabular}{c|c|c|c|c|c|c}
 		\toprule
 		Version&Round&Config&\textsf{Cor$_T$}&\textsf{Cor$_D$}&Gap&Source\\
 		\midrule
 		\multirow{4}{*}{\textsf{Simeck32}}
 		&12&$(5,2,5)$&$2^{-16.00}$&$2^{-9.00}$&$12700\%$&This work\\\cmidrule{2-7}
 		&\multirow{3}{*}{14}
 		&$(5,5,4)$&$2^{-16.63}$&$2^{-15.44}$&$128\%$&\cite{journals/iet/ZhouWH24}\\
 		&&$(5,6,3)$&\textcolor{blue}{$\bm{2^{-15.99}}$}&$2^{-15.45}$&$45\% $&This work\\
 		&&$(6,3,5)$&$2^{-20.00}$&\textcolor{red}{$\bm{2^{-14.73}}$}&$3758\%$&This work\\
 		\midrule
 		\multirow{6}{*}{\textsf{Simeck48}}
 		&\multirow{3}{*}{17}
 		&$(6,6,5)$&$2^{-22.37}$&$2^{-18.35}$&$1522\%$&\cite{journals/iet/ZhouWH24}\\
 		&&$(5,7,5)$&\textcolor{blue}{$\bm{2^{-21.44}}$}&$2^{-20.46}$&$97\%$&This work\\
 		&&$(8,4,5)$&$2^{-26.00}$&\textcolor{red}{$\bm{2^{-17.79}}$}&$29511\%$&This work\\\cmidrule{2-7}
 		&\multirow{3}{*}{18}
 		&$(6,6,6)$&$2^{-24.75}$&$2^{-20.80}$&$2445\%$&\cite{journals/iet/ZhouWH24}\\
 		&&$(8,7,3)$&\textcolor{blue}{$\bm{2^{-24.06}}$}&$2^{-20.70}$&$926\%$&This work\\
 		&&$(7,4,7)$&$2^{-28.00}$&\textcolor{red}{$\bm{2^{-20.41}}$}&$19167\%$&This work\\\cmidrule{2-7}
 		&19&$(8,4,7)$&$2^{-32.00}$&$2^{-23.22}$&$43858\%$&This work\\
 		\midrule
 		\multirow{12}{*}{\textsf{Simeck64}}
 		&\multirow{3}{*}{22}
 		&$(7,7,8)$&$2^{-32.44}$&$2^{-26.07}$&$8171\%$&\cite{journals/iet/ZhouWH24}\\
 		&&$(4,9,9)$&\textcolor{blue}{$\bm{2^{-29.04}}$}&$2^{-27.30}$&$234\%$&This work\\
 		&&$(9,6,7)$&$2^{-34.22}$&\textcolor{red}{$\bm{2^{-25.98}}$}&$30133\%$&This work\\\cmidrule{2-7}
 		&\multirow{3}{*}{23}
 		&$(7,7,9)$&$2^{-36.13}$&$2^{-31.06}$&$3259\%$&\cite{journals/iet/ZhouWH24}\\
 		&&$(4,10,9)$&\textcolor{blue}{$\bm{2^{-32.44}}$}&$2^{-30.98}$&$175\%$&This work\\
 		&&$(4,9,10)$&$2^{-33.15}$&\textcolor{red}{$\bm{2^{-30.92}}$}&$369\%$&This work\\\cmidrule{2-7}
 		&\multirow{3}{*}{24}
 		&$(7,7,10)$&$2^{-38.13}$&$2^{-32.35}$&$5394\%$&\cite{journals/iet/ZhouWH24}\\
 		&&$(4,9,11)$&\textcolor{blue}{$\bm{2^{-35.04}}$}&$2^{-33.31}$&$231\%$&This work\\
 		&&$(11,6,7)$&$2^{-40.00}$&\textcolor{red}{$\bm{2^{-31.88}}$}&$27720\%$&This work\\\cmidrule{2-7}
 		&\multirow{3}{*}{25}
 		&$(7,7,11)$&$2^{-41.04}$&$2^{-36.96}$&$1591\%$&\cite{journals/iet/ZhouWH24}\\
 		&&$(4,10,11)$&\textcolor{blue}{$\bm{2^{-39.04}}$}&$2^{-37.32}$&$229\%$&This work\\
 		&&$(11,7,7)$&$2^{-41.04}$&\textcolor{red}{$\bm{2^{-35.53}}$}&$4456\%$&This work\\
 		\bottomrule
 	\end{tabular}
 \end{table}


\subsection{Experimental Verification on Distinguishers}\label{Sect:experiments}
In this section, we perform some experiments on the obtained DL distinguishers to verify the validity of these distinguishers as well as the precision on the estimated correlations. For a given DL approximation with correlation $c$, it is sufficient to prepare $\frac{\epsilon}{c^2}$ pairs of chosen plaintext for distinguishing the target cipher from a pseudo-random permutation, where $\epsilon$ is a very small constant. Therefore, limited to the computational resource, we can only perform the experimental verification on the all distinguishers of \textsf{Simon32}/\textsf{Simeck32} and several high-correlation distinguishers of \textsf{Simon48}/\textsf{Simeck48} in practice. By preparing the appropriate number of plaintext samples, we can calculate the average of the absolute value of the correlation under 100 random master keys for a given DL distinguisher. The results are listed in Table~\ref{tab:experiments}. It can be seen that all the appearing distinguishers are valid and most of the estimated correlations are extremely close to the experimental ones, which proves the correctness and effectiveness of our improved automatic method especially the so-called transforming technique for finding DL distinguishers.

\begin{table}[!h]
	\centering
	\caption{Verification on some distinguishers by experiments. The number of plaintext samples for \textsf{Simon32}/\textsf{Simeck32} and \textsf{Simon48}/\textsf{Simeck48} are $2^{32}$ and $2^{34}$, respectively.}{\label{tab:experiments}}
	\renewcommand\tabcolsep{4pt}
	\begin{tabular}{c|c|c|c|c|c}
		\toprule
		Cipher & Round & \makecell{Input\\difference} & Output mask & \makecell{Estimated\\correlation} & \makecell{Experimental\\correlation} \\\midrule
		\multirow{8}{*}{\textsf{Simon32}} 
		&11&$(\mathtt{0x8},\mathtt{0x22})$&$(\mathtt{0x40},\mathtt{0x10})$&$2^{-8.03}$&$2^{-7.91}$\\\cmidrule{2-6}
		&\multirow{3}{*}{13} 
		&$(\mathtt{0x8},\mathtt{0x822})$&$(\mathtt{0x1000},\mathtt{0x4500})$&$2^{-13.92}$&$2^{-13.19}$\\
		&&$(\mathtt{0x8},\mathtt{0x22})$&$(\mathtt{0x40},\mathtt{0x110})$&$2^{-12.61}$&$2^{-10.91}$\\
		&&$(\mathtt{0x100, 0x440})$&$(\mathtt{0x800, 0x2200})$&$2^{-11.99}$&$2^{-10.95}$\\\cmidrule{2-6}
		&\multirow{3}{*}{14} 
		&$(\mathtt{0x8},\mathtt{0x22})$&$\mathtt{(0x400, 0x1101)}$&$2^{-15.63}$&$2^{-14.52}$\\
		&&$(\mathtt{0x1},\mathtt{0x4404})$&$(\mathtt{0x80},\mathtt{0x220})$&$2^{-16.75}$&$2^{-14.45}$\\
		&&$(\mathtt{0x100, 0x645})$&$(\mathtt{0x8000, 0x2002})$&$2^{-15.36}$&$2^{-14.48}$\\\cmidrule{2-6}
		&15&$(\mathtt{0x8,0x22})$&$(\mathtt{0x4,0x111})$&$2^{-18.29}$&$2^{-15.40}$\\\cmidrule{1-6}
		\multirow{3}{*}{\textsf{Simeck32}}
		&12&$(\mathtt{0x10,0x28})$&$(\mathtt{0x2,0x5})$&$2^{-9.00}$&$2^{-8.92}$\\\cmidrule{2-6}
		& \multirow{3}{*}{14}
		& $(\mathtt{0x100,0x2a0})$ & $(\mathtt{0x8,0x14})$ & $2^{-15.45}$ & $2^{-14.06}$   \\
		&& $(\mathtt{0x2000,0x7400})$ & $(\mathtt{0x100,0x288})$ & $2^{-15.44}$ & $2^{-14.07}$ \\
		&& $(\mathtt{0x4,0x800a})$ & $(\mathtt{0x2000,0x5000})$ &$2^{-14.73}$ & $2^{-14.04}$ \\\midrule

		\multirow{3}{*}{\textsf{Simon48}} &
		14&$(\mathtt{0x8,0x22})$ & $(\mathtt{0x80,0x220})$ & $2^{-13.33}$ & $2^{-11.51}$\\\cmidrule{2-6}
		&\multirow{3}{*}{15}
		&$(\mathtt{0x800,0x3200})$ & $(\mathtt{0x4,0x111})$ & $2^{-16.35}$ & $2^{-15.00}$\\
		&  & $(\mathtt{0x80},\mathtt{0x222})$ & $(\mathtt{0x20},\mathtt{0x88})$ & $2^{-16.83}$ & $2^{-15.09}$\\
		&& $(\mathtt{0x800,0x2220})$ & $(\mathtt{0x4,0x11})$ & $2^{-15.70}$ & $2^{-15.20}$\\
		\midrule
		\multirow{2}{*}{\textsf{Simeck48}} &
		\multirow{2}{*}{17} &
		$(\mathtt{0x800,0x1400})$&$(\mathtt{0x20,0x150})$&$2^{-18.35}$&$2^{-14.69}$\\
		&&($\mathtt{0x8000,0x15000}$)&$(\mathtt{0x200,0x500})$&$2^{-17.79}$&$2^{-14.71}$\\
		\bottomrule 
		
	\end{tabular}
\end{table}

By the way, Zhou \textit{et al.}~\cite{journals/iet/ZhouWH24} performed the \textit{segmented} experimental validations on the all presented DL trails. To be specific, they split a DL trail into two or three parts and tested the correlations for the two or three parts independently, and finally exploited the partial experimental correlations to compute the correlation of the entire DL trail according to piling-up lemma. For instance, they found the best 26-round DL trail with an estimated correlation of $2^{-50.66}$ under the configuration $(11,6,9)$. To verify this trail, they independently tested the 11-round differential part with an experimental probability of $2^{-26.60}$, the 6-round middle part with an experimental correlation of $2^{-0.49}$ and the 9-round linear part with an experimental squared correlation of $2^{-18.60}$. Merging the three parts by piling-up lemma, they deduced that the experimental correlation of this trail is about $2^{-45.69}$. Nevertheless, the segmented experiment based on piling-up lemma requires that the divided parts are independent, which can hardly be guaranteed in practice. Therefore, these experimental correlations computed in this way are not convincing enough. The details can be referred to the appendix in~\cite{journals/iet/ZhouWH24}.   

\section{Conclusion}\label{Sect:5}
In this paper, we enhanced the automatic method for finding DL distinguishers of \textsf{Simon}-like block ciphers from three perspectives: 1) constructing the model for finding the best DL trail with precision and ease; 2) speeding up the search process by the proposed differential-first and linear-first heuristic strategies; 3) introducing the transforming technique to exploit the clustering effect on DL trails. With the improved method, we found the best DL distinguishers for the all variants of \textsf{Simon} block cipher family so far. It is worth noting that the transforming technique is a general approach, thus it is applicable to other works about searching for DL distinguishers based on the three-part structure by automatic tools, like~\cite{DBLP:conf/ctrsa/BelliniGGMP23} and~\cite{DBLP:journals/iacr/HadipourDE24}, which may lead the better results.

	
During the process of searching for DL trails and transforming them to DL distinguishers, we have found a interesting phenomenon as mentioned in Section~\ref{Sect:4.2}. That is, for a fixed entire round, a DL trail with the round configuration where the middle part is longer (resp. shorter) but the differential/linear part is shorter (resp. longer), always has the two features: 1) it has the higher (resp. lower) correlation; 2) it leads a DL distinguisher with the lower (resp. higher) correlation. The reason for the first feature, we guess, is that the correlation of middle part will decrease slowly but the probability/correlation of differential/linear part will drop rapidly with the round increasing. For the second feature, we guess the cause possibly comes from the imprecise estimated correlation on the continuous difference propagation in the middle part. Because the precondition that the propagation rules of continuous difference through XOR and AND operations hold is all the internal bits are independent of each other, which is not consistent with the practical situation. Thus, the shorter the middle part is, the closer its estimated correlation will be to the exact one. Besides, the clustering effect will be more and more significant with the round of differential/linear part increasing. This may also be the reason why the estimated correlation of the DL distinguisher that has a configuration with the shorter middle part is much closer to the experimental one. Therefore, if we can describe the middle part in a more precise way, then the estimated correlation will be further improved. A potential approach that may make sense is to investigate the continuous difference propagation through the round function based on the work of K\"{o}lbl \textit{et al.}~\cite{DBLP:conf/crypto/KolblLT15}, which is worthy of a deep studying in the future.  

 
	
\newcommand{\etalchar}[1]{$^{#1}$}

\newpage
\appendix
\section{The Detail of Modeling the Differential Part}\label{appendix:A}
We introduce the $n$-bit variables $\alpha,\beta,\gamma,\texttt{varibits},\texttt{doublebits}$ corresponding Theorem~\ref{thm:simon_pro}. Thus the relation $\texttt{varibits}=S^a(\alpha)\vee S^b(\alpha)$ can be constrained by
\begin{equation}\label{equ:model_varibits}
	\begin{cases}
		\texttt{varibits}_i - \alpha_{(i+a)\text{mod}n} \geq 0\\
		\texttt{varibits}_i - \alpha_{(i+b)\text{mod}n} \geq 0\\
		\alpha_{(i+a)\text{mod}n} + \alpha_{(i+b)\text{mod}n} - \texttt{varibits}_i \geq 0\\
		\text{for each }i\in\{0,1,...,n-1\}.
	\end{cases}
\end{equation}
For the sake of simplicity, we straightforwardly re-write the index $(i+j)\text{mod}n$ as $i+j$ here as well as in the modeling processes of other two parts. Similarly, we use the linear inequalities
\begin{equation}\label{equ:model_doublebits}
	\begin{cases}
		\texttt{doublebits}_i + \alpha_{i+a} \leq 1\\
		\texttt{doublebits}_i - \alpha_{i+b} \leq 0\\
		\texttt{doublebits}_i - \alpha_{i+2a-b} \leq 0\\
		\texttt{doublebits}_i + \alpha_{i+a} - \alpha_{i+b} - \alpha_{i+2a-b} \geq -1\\
		\text{for each }i\in\{0,1,...,n-1\}.
	\end{cases}
\end{equation}
to describe the relation $\texttt{doublebits}=S^b(\alpha)\wedge\overline{S^a(\alpha)}\wedge S^{2a-b}(\alpha)$. The output difference of round function $\beta = \gamma \oplus S^c(\alpha)$ is described as
\begin{equation}\label{equ:model_beta}
	\begin{cases}
		\gamma_i + \alpha_{i+c} - \beta_i \geq 0\\
		\gamma_i - \alpha_{i+c} + \beta_i \geq 0\\
		-\gamma_i + \alpha_{i+c} + \beta_i \geq 0\\
		\gamma_i + \alpha_{i+c} + \beta_i \leq 2\\
		\text{for each }i\in\{0,1,...,n-1\}.
	\end{cases}
\end{equation}
Note that we always expect the probability can be high as much as possible, which requires that the case of $\alpha=1^n$ does never appear. Thus we need to constrain the weight of $\alpha$ being less than $n$, i.e., 
\begin{equation}\label{equ:model_alpha}
	\sum_{i=0}^{n-1}\alpha_i \leq n-1.
\end{equation}
For the case of $\alpha\neq 1^n$, the conditions that $\gamma$ needs to satisfy $\gamma \wedge \overline{\texttt{varibits}}=0^n$ and $(\gamma\oplus S^{a-b}(\gamma))\wedge \texttt{doublebits}=0^n$, can be constrained by
\begin{equation}\label{equ:model_gamma}
	\begin{cases}
		\gamma_i - \texttt{varibits}_i \leq 0\\
		\gamma_i - \gamma_{i+a-b} + \texttt{doublebits}_i \leq 1\\ 
		-\gamma_i + \gamma_{i+a-b} + \texttt{doublebits}_i \leq 1\\
		\text{for each }i\in\{0,1,...,n-1\}.
	\end{cases}
\end{equation}
We use an $n$-bit auxiliary variable $\texttt{probits}$ to represent $\texttt{variabits}\oplus\texttt{doublebits}$ and an auxiliary integer variable $\texttt{pro}$ to represent the weight of $\texttt{probits}$, then we have
\begin{equation}\label{equ:model_pro}
	\begin{cases}
		\texttt{varibits}_i + \texttt{doublebits}_i - \texttt{probits}_i \geq 0\\
		\texttt{varibits}_i - \texttt{doublebits}_i + \texttt{probits}_i \geq 0\\
		-\texttt{varibits}_i + \texttt{doublebits}_i + \texttt{probits}_i \geq 0\\
		\texttt{varibits}_i + \texttt{doublebits}_i + \texttt{probits}_i \leq 2\\
		\text{for each }i\in\{0,1,...,n-1\}\\
		\texttt{pro} - \sum_{j=0}^{n-1}\texttt{probits}_j = 0.
	\end{cases}
\end{equation}
In summary, the combination of Equation~\eqref{equ:model_varibits}-\eqref{equ:model_pro}, which contains $(18n+2)$ purely linear constraints and $(6n+1)$ mixed-integer variables, can be used to model the exact relation between input and output differences of the round function as well as the probability, i.e., $2^{-\texttt{pro}}$. 

\qquad\qquad
\section{The Piece-wise Linear Function in~\cite{DBLP:conf/ctrsa/BelliniGGMP23}}\label{appendix:B}
The piece-wise linear function $g(x)$ (Equation~\eqref{equ:model-abs-log2}) given by Bellini \textit{et al.} to approximate $-\log_2|x|$. 
\begin{equation}\label{equ:model-abs-log2}
g(x) = 
\begin{cases}
	-19931.570x+29.897, & 0\leq x\leq0.001\\
	-584.962x + 10.135, & 0.001\leq x \leq 0.004\\
	-192.645x + 8.506, & 0.004\leq x \leq 0.014\\	
	-50.626x + 6.575, & 0.014\leq x \leq 0.053\\
	-11.87x + 4.483, & 0.053\leq x \leq 0.142\\
	-8.613x + 4.020, & 0.142\leq x \leq 0.246\\
	-3.761x + 2.825, & 0.246 \leq x \leq 0.595\\
	-1.444x + 1.444, & 0.595 \leq x \leq 0.998.
\end{cases}
\end{equation}

\newpage
\section{Algorithms Mentioned in Section~\ref{Sect:3}}\label{appendix:algorithms}

\begin{algorithm}[!h]\label{alg:model-diff-part}
	\caption{Modeling the differential part of \textsf{Simon}-like ciphers}
	\SetAlgoLined
	\KwIn{The round $R_d$.}
	\KwOut{An model $\mathcal{M}_d$; The logarithm on differential propbability denoted by $\mathtt{Pro}$; The left and right branches of output difference denoted by the $n$-bit variables $\alpha^{R_d+1}$ and $\alpha^{R_d}$, respelctively.}
	\SetKwFunction{ModelDiffPart}{ModelDiffPart} 
	\SetKwProg{Fn}{Function}{:}{\KwRet{$(\mathcal{M}_d, \mathtt{Pro}, \alpha^{R_d+1}, \alpha^{R_d})$. }} 
	\Fn{\ModelDiffPart{$R_d$}}{
		Prepare an empty model $\mathcal{M}_d$;\\
		Declare two $n$-bit variables $\alpha^0=(\alpha^0_0,...,\alpha^0_{n-1})$ and $\alpha^1=(\alpha^1_0,...,\alpha^1_{n-1})$; \textcolor{blue}{{\# The right and left branches of initial difference.}}\\
		
		Add constraint $\sum_{i=0}^{n-1}(\alpha^0_i+\alpha^1_i)\geq1$ into $\mathcal{M}_d$; \textcolor{blue}{{\# Let the input difference be non-trivial.}}\\
		\For{$r$ from $0$ to $R_d-1$}{
			Declare an $n$-bit variable $\alpha^{r+2}$; \textcolor{blue}{\# {The left branch of next round difference.}}\\ 
			Declare four $n$-bit variables $\beta^r$, $\gamma^r$, $\mathtt{varibits}^r$ and $\mathtt{doublebits}^r$;\\
			Declare an $n$-bit variable $\mathtt{probits}^r$;\\
			Declare a variable $\mathtt{pro}^r\in\mathbb{Z}$; \textcolor{blue}{{\# The $r$-round probability is $2^{-\mathtt{pro}^r}$.}}\\
			Add constraints from Equation~\eqref{equ:model_varibits}-\eqref{equ:model_pro} into $\mathcal{M}_d$; \textcolor{blue}{{\# Modeling $\alpha^{r+1}$, $\beta^r$ and $\mathtt{pro}^r$.}}\\
			Add constraints that describe $\beta^r \oplus \alpha^r = \alpha^{r+2}$ into $\mathcal{M}_d$; \textcolor{blue}{{\# Similarly to Equation~\eqref{equ:model_beta}.}}\\
		}
		Declare a variable $\mathtt{Pro}\in\mathbb{Z}$ and add the constraint $\mathtt{Pro} = \sum_{r=0}^{R_d-1}\mathtt{pro}^r$ into $\mathcal{M}_d$;\\
	}
\end{algorithm}

\begin{algorithm}[!h]\label{alg:model-middle-part}
	\caption{Modeling the middle part of \textsc{Simon}-like ciphers}
	\SetAlgoLined
	\KwIn{The round $R_m$; The output difference $(\alpha^{R_d+1}, \alpha^{R_d})$ of $R_d$-round differential part; The initial mask $(\lambda^0,\lambda^1)$ of the linear part.}
	\KwOut{An model $\mathcal{M}_m$; The logarithm on the absolute value of correlation denoted by $\mathtt{Cor}_m$.}
	\SetKwFunction{ModelMiddlePart}{ModelMiddlePart} 
	\SetKwProg{Fn}{Function}{:}{\KwRet{$(\mathcal{M}_m, \mathtt{Cor}_m)$.}} 
	\Fn{\ModelMiddlePart{$R_m$, $\alpha^{R_d+1}$, $\alpha^{R_d}$, $\lambda^0$, $\lambda^1$}}{
		Prepare an empty model $\mathcal{M}_m$;\\
		Declare two vectorial variable $x^0=(x^0_0,...,x^0_{n-1})\in\mathbb{R}^n$ and $x^1=(x^1_0,...,x^1_{n-1})\in\mathbb{R}^n$; \textcolor{blue}{{\# The right and left branches of initial continuous difference.}}\\
		Add constraints from Equation~\eqref{equ:model_initial_CD}
		into $\mathcal{M}_m$;\\
		\For{$r$ from $0$ to $R_m-1$}{
			Declare three vectorial variables $x^{r+2}$, $y^{r}$, $t^{r}$ that belong to $\mathbb{R}^n$;\\ 
			Add constraints from Equation~\eqref{equ:model_AND-CD} and~\eqref{equ:model_XOR-CD} into $\mathcal{M}_m$;\\
		}
		Declare four vectorial variables $z^i\in\mathbb{R}^n(0\leq i\leq3)$ and a variable $\mathtt{Cor}_m\in\mathbb{R}$;\\
		Add constraints from Equation~\eqref{equ:model_middle_Cor} into $\mathcal{M}_m$;
	}
\end{algorithm}

\begin{algorithm}[!h]\label{alg:calculate-cor}
	\caption{Calculating the correlation of \textsf {Simon}-like round function}
	\SetAlgoLined
	\KwIn{The concrete input mask ($\lambda^0,\lambda^1$) and the correponding output mask ($\lambda^1,\lambda^2$) of the one-round encryption.}
	\KwOut{The correlation of the mask ($\lambda^0,\lambda^1$) propagating to ($\lambda^1,\lambda^2$).}
	
	$\lambda_{out} = \lambda^1$;\\
	$\lambda_{in} = \lambda^0\oplus\lambda^2\oplus S^{-c}(\lambda^1);$\\
	\If{$((S^{-a}(\lambda_{out})\vee S^{-b}(\lambda_{out})) \oplus \lambda_{in})\wedge\lambda_{in} \ \neq \  0$}{
		\textbf{return} 0;
	}
	\If{$\lambda_{out} == 1^n$}{
		$v = 0^n$;\\
		\While{$\lambda_{in}\neq 0^n$}{
			$v = v\oplus \lambda_{in} \wedge 0x3$;\\
			$\lambda_{in} = \lambda_{in}/4$;\\
			\eIf{$v\neq 0^n$}
			{
				\textbf{return} 0;
			}
			{
				\textbf{return} $2^{-n+2}$;
			}
		} 
		
	}
	$\mathtt{(tmp, abits)}=(\lambda_{out},\lambda_{out})$;\\
	\While{$\mathtt{tmp}\neq0^n$}{
		$\mathtt{tmp}=\lambda_{out}\wedge S^{b-a}(\mathtt{tmp})$;\\
		$\mathtt{abits}=\mathtt{abits}\oplus\mathtt{tmp}$;	
	}
	$\mathtt{sbits}=S^{-a}(\lambda_{out})\wedge\overline{S^{-b}(\lambda_{out})}\wedge\overline{S^{-a}(\mathtt{abits})}$;\\
	$\mathtt{pbits}=S^{2a-2b}(\mathtt{
		sbits}\wedge\lambda_{in})$;\\
	\While{$\mathtt{sbits}\neq0^n$}{
		$\mathtt{sbits}=S^{2a-2b}(\mathtt{sbits})\wedge S^{a-2b}(\lambda_{out})$;\\
		$\mathtt{pbits} = S^{2a-2b}(\mathtt{
			sbits}\wedge\lambda_{in}\oplus\mathtt{pbits})$;\\
	}
	\eIf{$\mathtt{pbits}\neq 0^n$}{
		\textbf{return} 0;
	}
	{
		\textbf{return} $2^{-wt(\mathtt{abits})}$.
	}
\end{algorithm}

\begin{algorithm}[H]\label{alg:model-linear-part}
	\caption{Modeling the linear part of \textsc{Simon}-like ciphers}
	\SetAlgoLined
	\KwIn{The round $R_l$.}
	\KwOut{An model $\mathcal{M}_l$; The logarithm on the absolute value of correlation denoted by $\mathtt{Cor}_l$; The left and right branches of input mask denoted by the $n$-bit variables $\lambda^{0}$ and $\lambda^{1}$, respelctively.}
	\SetKwFunction{ModelLinearPart}{ModelLinearPart} 
	\SetKwProg{Fn}{Function}{:}{\KwRet{$(\mathcal{M}_l, \mathtt{Cor}_l, \lambda^{0},\lambda^{1})$.}} 
	\Fn{\ModelLinearPart{$R_l$}}{
		Prepare an empty model $\mathcal{M}_l$;\\
		Declare two $n$-bit variables $\lambda^0=(\lambda^0_0,...,\lambda^0_{n-1})$ and $\lambda^1=(\lambda^1_0,...,\lambda^1_{n-1})$; \textcolor{blue}{{\# The left and right branches of initial mask.}}\\
		Add constraint $\sum_{i=0}^{n-1}(\lambda^0_i+\lambda^1_i)\geq1$ into $\mathcal{M}_l$; \textcolor{blue}{{\# Let the input mask be non-trivial.}}\\
		\For{$r$ from $0$ to $R_l-1$}{
			Declare two $n$-bit variables $\lambda^{r+2}$ and $\mathtt{abits}^r$;\\ 
			Declare $n$ vectorial variables $\mathtt{tmp}^{0,r},...,\mathtt{tmp}^{n-1,r}$ that belong to $\mathbb{F}_2^n$;\\
			Declare a vectorial variable $N^r\in\mathbb{Z}^n$;\\
			Declare $n$ vectorial variables $\mathtt{sbits}^{0,r},...,\mathtt{sbits}^{n-1,r}$ that belong to $\mathbb{F}_2^n$;\\
			Declare $n$ vectorial variables $\mathtt{pbits}^{0,r},...,\mathtt{pbits}^{n-1,r}$ that belong to $\mathbb{F}_2^n$;\\
			Declare a variable $\mathtt{Cor}^r\in\mathbb{Z}$;\\
			Add constraints from Equation~\eqref{equ:model-lambda_in}-\eqref{equ:model-linear-cor} into $\mathcal{M}_l$;\\
		}
		Declare a variable $\mathtt{Cor}_l\in\mathbb{Z}$ and add constraint $\mathtt{Cor}_l=\sum_{r=0}^{R_l-1}\texttt{Cor}^r$ into $\mathcal{M}_l$;
	}
\end{algorithm}

\begin{algorithm}[!h]\label{alg:model-entire}
	\caption{Finding the $R$-round best DL trail with a given round configuration}
	\SetAlgoLined
	\KwIn{The round configuration $(R_d,R_m,R_l)$ satisfying $R=R_d+R_m+R_l$.}
	\KwOut{An $R$-round DL trail.}
	$(\mathcal{M}_d, \mathtt{Pro},\alpha^{R_d+1},\alpha^{R_d})=$ \texttt{ModelDiffPart(}$R_d$\texttt{)};\\
	$(\mathcal{M}_l,\mathtt{Cor}_l,\lambda^{0},\lambda^{1})=$ \texttt{ModelLinearPart(}$R_l$\texttt{)};\\
	$(\mathcal{M}_m,\mathtt{Cor}_m)=$ \texttt{ModelMiddlePart(}$R_m,\alpha^{R_d+1},\alpha^{R_d},\lambda^0,\lambda^1$\texttt{)};\\
	Construct $\mathcal{M}_e$ by merging $\mathcal{M}_d,\mathcal{M}_m$ and $\mathcal{M}_l$;\\
	Declare a variable $\mathtt{Cor}_e\in\mathbb{R}$; \textcolor{blue}{ \# The correlation of the DL approximation.}\\
	Add constraint $\mathtt{Cor}_e = \mathtt{Pro} - \mathtt{Cor}_m + 2\cdot\mathtt{Cor}_l$ into $\mathcal{M}_e$;\\
	$\mathcal{M}_e.\textsf{setObjective}(\mathtt{Cor}_e, \mathsf{sense=GRB.MINIMIZE})$; \textcolor{blue}{ \# \textsf{Python} interface of \textsf{Gurobi}: set the objective function.}\\
	$\mathcal{M}_e.\textsf{optimize()}$; \textcolor{blue}{ \# \textsf{Python} interface of \textsf{Gurobi}: solve the model $\mathcal{M}_e$.}\\
	Retrieve the values of $\mathtt{Cor}_e, \mathtt{Pro},\mathtt{Cor}_l,\mathtt{Cor}_m,(\alpha^1,\alpha^0), (\alpha^{R_d+1},\alpha^{R_d}),(\lambda^0,\lambda^1)$, $(\lambda^{R_l},\lambda^{R_l+1})$ from the solution returned by \textsf{Gurobi}.
\end{algorithm}

\begin{algorithm}[!h]\label{alg:calculate-continuous-diff}
	\caption{Calculating the continuous difference from a given difference}
	\SetAlgoLined
	\KwIn{The round $R_m$; The concrete input difference $(\delta^L,\delta^R)\in\mathbb{F}_2^{n}\times\mathbb{F}_2^n$.}
	\KwOut{The concrete output continuous difference $(x^L,x^R)$.}
	\SetKwFunction{ComputeContinDiff}{ComputeContinDiff} 
	\SetKwProg{Fn}{Function}{:}{\KwRet{$(x^{R},x^{L})$.}} 
	\Fn{\ComputeContinDiff{$R_m,\delta^L,\delta^R$}}{
		Initial both $x^{L}\in\mathbb{R}^n$ and $x^{R}\in\mathbb{R}^n$ as $(0,...,0)$;\\
		\For{i from 0 to $n-1$}
		{
			$x^L_i=1-2\cdot\delta^L_i$;\\
			$x^R_i=1-2\cdot\delta^R_i$;\\
		}
		\For{$r$ from $0$ to $R_m-1$}{
			Initial $tmp\in\mathbb{R}^n$ as $(0,...,0)$;\\
			\For{i from 0 to n}
			{
				$tmp_i = x^L_i$;\\
				$x^L_i = 0.25\cdot(1+x^{L}_{(i+a)\%n}+x^{L}_{(i+b)\%n}+x^{L}_{(i+a)\%n}\cdot x^{L}_{(i+b)\%n})\cdot x^{L}_{i+c}\cdot x^{R}_i;$\\
				$x^{R}_i =  tmp_i$;\\
			}
		}
	}
\end{algorithm}
\begin{algorithm}[!h]\label{alg:differential-first-strategy}
	\caption{Finding an $R$-round DL trail with the differential-first strategy}
	\SetAlgoLined
	\KwIn{The round configuration $(R_d,R_m,R_l)$ satisfying $R=R_d+R_m+R_l$.}
	\KwOut{An $R$-round DL trail.}
	$(\mathcal{M}_d, \mathtt{Pro},\alpha^{R_d+1},\alpha^{R_d})=$ \ModelDiffPart{$R_d$};\\
	$\mathcal{M}_d.\textsf{setObjective}(\mathtt{Pro}, \mathsf{sense=GRB.MINIMIZE})$;\\
	$\mathcal{M}_d.\textsf{optimize()}$;\\
	Retrieve the values of $ \mathtt{Pro},(\alpha^1,\alpha^0), (\alpha^{R_d+1},\alpha^{R_d})$ from the solution returned by \textsf{Gurobi}; \textcolor{blue}{ \# {An $R_d$-round best differential characteristic}.}\\
	$(x^{L},x^{R})=$ \ComputeContinDiff{$R_m,\alpha^{R_d+1}, \alpha^{R_d}$};\\
	$x = x^{L}||x^{R}$;\\
	\For{i from 0 to $2n-1$}
	{

		\eIf{$x_i==0$}
			{$x_i=-n^2$;}
			{$x_i=\log_2|x_i|$;}

	}
	$x^L||x^R = x$;\\
	$(\mathcal{M}_l,\mathtt{Cor}_l,\lambda^0,\lambda^1)=$ \ModelLinearPart{$R_l$};\\
	Declare two variable $\mathtt{Cor}_e\in\mathbb{R}$ and $\mathtt{Cor}_m\in\mathbb{R}$;\\
	Add constraint $\mathtt{Cor}_m = \sum_{i=0}^{n-1}(x^{L}_i\cdot\lambda^0_i + x^{R}_i\cdot\lambda^1_i)$ into $\mathcal{M}_l$;  \textcolor{blue}{ \# {The values of $x^{L}_i$ and $x^{R}_i$ are known.}}\\
	Add constraint $\mathtt{Cor}_e = \mathtt{Pro} - \mathtt{Cor}_m + 2\cdot\mathtt{Cor}_l$ into $\mathcal{M}_l$; \textcolor{blue}{ \# {The value of $\mathtt{Pro}$ is known.}}\\
	$\mathcal{M}_l.\textsf{setObjective}(\mathtt{Cor}_e, \mathsf{sense=GRB.MINIMIZE})$;\\
	$\mathcal{M}_l.\textsf{optimize()}$;\\
	Retrieve the values of $ \mathtt{Cor}_e,\mathtt{Cor}_m,\mathtt{Cor}_l, (\lambda^0,\lambda^1), (\lambda^{R_l},\lambda^{R_l+1})$ from the solution returned by \textsf{Gurobi}.\\
\end{algorithm}

\begin{algorithm}[H]\label{alg:linear-first-strategy}
	\caption{Finding an $R$-round DL trail with the linear-first strategy}
	\SetAlgoLined
	\KwIn{The round configuration $(R_d,R_m,R_l)$ satisfying $R=R_d+R_m+R_l$.}
	\KwOut{An $R$-round DL trail.}
	$(\mathcal{M}_l, \mathtt{Cor}_l,\lambda^{0},\lambda^{1})=$ \ModelLinearPart{$R_l$};\\
	$\mathcal{M}_l.\textsf{setObjective}(\mathtt{Cor}_l, \mathsf{sense=GRB.MINIMIZE})$;\\
	$\mathcal{M}_l.\textsf{optimize()}$;\\
	Retrieve the values of $ \mathtt{Cor}_l,(\lambda^0,\lambda^1), (\lambda^{R_l},\lambda^{R_l+1})$ from the solution returned by \textsf{Gurobi}.	\textcolor{blue}{ \# {An $R_l$-round best linear trail}.}\\
	$(\mathcal{M}_d, \mathtt{Pro},\alpha^{R_d+1},\alpha^{R_d})=$ \ModelDiffPart{$R_d$};\\
	$(\mathcal{M}_m,\mathtt{Cor}_m)=$ \ModelMiddlePart{$R_m,\alpha^{R_d+1},\alpha^{R_d},\lambda^0,\lambda^1$};\\
	Merge $\mathcal{M}_d$ and $\mathcal{M}_m$ to a new model $\mathcal{M}_e$;\\
	Declare a variable $\mathtt{Cor}_e\in\mathbb{R}$;\\
	Add constraint $\mathtt{Cor}_e=\mathtt{Pro}-\mathtt{Cor}_m+2\cdot\mathtt{Cor}_l$ into $\mathcal{M}_e$; \textcolor{blue}{ \# {The value of $\mathtt{Cor}_l$ is known.}}\\
	$\mathcal{M}'_e.\textsf{setObjective}(\mathtt{Cor}_e, \mathsf{sense=GRB.MINIMIZE})$;\\
	$\mathcal{M}'_e.\textsf{optimize()}$;\\
	Retrieve the values of $ \mathtt{Cor}_e,\mathtt{Cor}_m,\mathtt{Pro}, (\alpha^1,\alpha^0), (\alpha^{R_d+1},\alpha^{R_d})$ from the solution returned by \textsf{Gurobi}.\\
\end{algorithm}
\newpage
\begin{algorithm}[H]\label{alg:transforming-technique}
	\caption{Transforming a DL trail to a DL distinguisher}
	\SetAlgoLined
	\KwIn{A DL trail contains: configuration $(R_d,R_m,R_l)$, difference-mask pair $(\Delta_I,\lambda_O)$, differential probability $p$, and linear approximation $q$; The chosen lower bound on differential probability and linear approximation: $\overline{p}$ and $\overline{q}$.}
	\KwOut{An DL distinguisher with a newly estimated correlation $\mathsf{Cor}_D$.}
	Prepare two empty lists $\mathbb{L}_d$ and $\mathbb{L}_l$;\\ $(\mathcal{M}_d,\mathtt{Pro},\alpha^{R_d+1},\alpha^{R_d})=$\ModelDiffPart{$R_d$};\\
	Add inequalities that constrain $\alpha^1||\alpha^0=\Delta_I$ into $\mathcal{M}_d$;\\
	\For{P from $-\log_2p$ to $-\log_2\overline{p}$}
	{    
		Construct a new model $\mathcal{M}$ copied from $\mathcal{M}_d$;\\
		Add constraint $\mathtt{Pro}=P$ into $\mathcal{M}$;\\
		\While{$\mathcal{M}$ \text{is solvable}}
		{
			Retrieve the value of $\alpha^{R_d+1}||\alpha^{R_d}$ that denoted by $\Delta=\delta^L||\delta^R$;\\ 
			Add $(2^{-P},\Delta)$ into $\mathbb{L}_d$;\\
			Add constrain $\sum_{i=0}^{n-1}[
			(-1)^{\delta^L_i}\cdot\alpha^{R_d+1}_i + (-1)^{\delta^R_i}\cdot\alpha^{R_d}_i + \delta^L_i+\delta^R_i ] \geq 1$ into $\mathcal{M}$; \textcolor{blue}{ \# {Remove this output difference from the solution space of $\mathcal{M}$.}}\\
			
		}
	}
	$(\mathcal{M}_l, \mathtt{Cor}_l,\lambda^{0},\lambda^{1})=$ \ModelLinearPart{$R_l$};\\
	Add inequalities that constrain $\lambda^{R_l}||\lambda^{R_l+1}=\lambda_O$ into $\mathcal{M}_l$;\\
	\For{Q from $-\log_2q$ to $-\log_2\overline{q}$}
	{    
		Construct a new model $\mathcal{M}'$ copied from $\mathcal{M}_l$;\\
		Add constraint $\mathtt{Cor}_l=Q$ into $\mathcal{M}'$;\\
		\While{$\mathcal{M}'$ \text{is solvable}}
		{
			Retrieve the value of $\lambda^{0}||\lambda^{1}$ that denoted by $\lambda=\eta^L||\eta^R$;\\ 
			Add $(2^{-Q},\lambda)$ into $\mathbb{L}_l$;\\
			Add constrain $\sum_{i=0}^{n-1}[
			(-1)^{\eta^L_i}\cdot\lambda^{0}_i + (-1)^{\eta^R_i}\cdot\lambda^{1}_i + \eta^L_i+\eta^R_i ] \geq 1$ into $\mathcal{M}'$; \textcolor{blue}{ \# {Remove this input mask from the solution space of $\mathcal{M}'$.}}\\
			
		}
	}
	Initial $\textsf{Cor}_D = 0$;\\
	\For{each pair $(p_*,\Delta)$ in $\mathbb{L}_d$}
	{
		
		$\delta^L||\delta^R = \Delta$;\\
		$(x^L,x^R) =$ \ComputeContinDiff{$R_m,\delta^L,\delta^R$};\\
		$x = x^L||x^R$;\\
		\For{each pair $(q_*,\lambda)$ in $\mathbb{L}_l$}
		{
			Initial $tmp=1$;\\
			\For{$i$ from $0$ to $2n-1$}
			{
				\If{$\lambda_i == 1$}{
					$tmp = tmp \cdot x_i$;\\
				}
			}
			$\mathsf{Cor}_D = \mathsf{Cor}_D + p_* \cdot tmp \cdot  q_*^{2}$;\\
		}
	}
	\textbf{return} $\textsf{Cor}_D$.
\end{algorithm}

\newpage
\section{The Details of DL Trails in Section~\ref{Sect:4.1} and~\ref{Sect:4.2}}\label{appendix:DL-trails}	

This part gives the specification on the DL trails that appear in Table~\ref{tab:DL-trails},~\ref{tab:best-DL-distinguisher},~\ref{tab:DL-distinguisher-simon} and~\ref{tab:DL-distinguisher-simeck} in Section~\ref{Sect:4.1} and~\ref{Sect:4.2}. In the following tables, the first column indicates the round and configuration. Also, the round number marked with `*' means it is a new trail found in our work, on the contrary, it is presented in~\cite{journals/iet/ZhouWH24}. The second and third columns represent the differential and linear characteristics of $E_d$ and $E_l$, respectively. By the way, the correlations of $E_m$ and $E$ listed here, i.e. $\textsf{Cor}_{E_m}$ and $\textsf{Cor}_E$, are the absolute values.

 \begin{table}[!h]
	\centering
	\caption{The details of DL trails for $\mathsf{Simon32}$.}{\label{tab:DL-trail-simon32}}

	
			\begin{tabular}{ccccccc}
				\toprule
				Round&$\Delta_I\overset{E_d}{\rightarrow}\Delta$&$\lambda\overset{E_l}{\rightarrow}\lambda_O$&$\textsf{Pro}_{E_d}$&$\mathsf{Cor}_{E_m}$&$\mathsf{Cor}^2_{E_l}$&$\mathsf{Cor}_E$\\
				\midrule
				\makecell{11*\\$(5,2,4)$}&\makecell{$(\mathtt{0x8,0x22 })$\\$(\mathtt{0x22,0x8   })$}&\makecell{$(\mathtt{0x44,0x10})$\\$(\mathtt{0x40,0x10})$}&$2^{-8}$&$1$&$2^{-6}$&$2^{-14}$\\
				\midrule
				\makecell{13\\$(5,5,3)$}&\makecell{$(\mathtt{0x800},\mathtt{0x2208})$\\$(\mathtt{0x2200},\mathtt{0x800})$}&\makecell{$(\mathtt{0x0},\mathtt{0x100})$\\$(\mathtt{0x10},\mathtt{0x45})$}&$2^{-8}$&$2^{-0.63}$&$2^{-8}$&$2^{-16.63}$\\
				\midrule
				\makecell{13*\\$(5,5,3)$}&\makecell{$(\mathtt{0x8},\mathtt{0x22})$\\$(\mathtt{0x22},\mathtt{0x8})$}&\makecell{$(\mathtt{0x100},\mathtt{0x0})$\\$(\mathtt{0x40},\mathtt{0x110})$}&$2^{-8}$&$2^{-2.73}$&$2^{-4}$&$2^{-14.73}$\\
				\midrule
				\makecell{13*\\$(5,3,5)$}&\makecell{$(\mathtt{0x100,0x440})$\\$\mathtt(0x440,0x100)$}&\makecell{$\mathtt{(0x2208,0x800)}$\\$(\mathtt{0x800,0x2200})$}&$2^{-8}$&$2^{-0.83}$&$2^{-8}$&$2^{-16.83}$\\
				\midrule
				\makecell{14\\$(5,5,4)$}&\makecell{$(\mathtt{0x8},\mathtt{0x22})$\\$(\mathtt{0x22},\mathtt{0x8})$}&\makecell{$\mathtt{(0x0, 0x1)}$\\$\mathtt{(0x400, 0x1101)}$}&$2^{-8}$&$2^{-0.63}$&$2^{-10}$&$2^{-18.63}$\\
				\midrule
				\makecell{14*\\$(4,6,4)$}&\makecell{$(\mathtt{0x4},\mathtt{0x11})$\\$(\mathtt{0x4},\mathtt{0x1})$}&\makecell{$\mathtt{(0x0, 0x8000)}$\\$\mathtt{(0x200, 0x8880)}$}&$2^{-6}$&$2^{-1.88}$&$2^{-10}$&$2^{-17.88}$\\
				\midrule
				\makecell{14*\\$(7,3,4)$}&\makecell{$(\mathtt{0x100},\mathtt{0x645})$\\$(\mathtt{0x44},\mathtt{0x10})$}&\makecell{$\mathtt{(0x8000, 0x2)}$\\$\mathtt{(0x8000, 0x2002)}$}&$2^{-14}$&$1$&$2^{-6}$&$2^{-20}$\\
				\midrule
				\makecell{15*\\$(5,5,5)$}&\makecell{$(\mathtt{0x80},\mathtt{0x220})$\\$(\mathtt{0x220},\mathtt{0x80})$}&\makecell{$\mathtt{(0x0, 0x1000)}$\\$\mathtt{(0x40, 0x1110)}$}&$2^{-8}$&$2^{-2.73}$&$2^{-10}$&$2^{-20.73}$\\
				\bottomrule
			\end{tabular}

\end{table}

\begin{table}[!h]
	\centering
	\caption{The details of DL trails for $\mathsf{Simon48}$.}{\label{tab:DL-trail-simon48}}
	\renewcommand\tabcolsep{5pt}
	\begin{threeparttable}
			\begin{tabular}{ccccccc}
				\toprule
				Round&$\Delta_I\overset{E_d}{\rightarrow}\Delta$&$\lambda\overset{E_l}{\rightarrow}\lambda_O$&$\textsf{Pro}_{E_d}$&$\mathsf{Cor}_{E_m}$&$\mathsf{Cor}^2_{E_l}$&$\mathsf{Cor}_E$\\
				\midrule
				\makecell{14*\\$(5,4,5)$}&\makecell{$(\mathtt{0x8,0x22 })$\\$(\mathtt{0x22,0x8})$}&\makecell{$(\mathtt{0x220,0x80})$\\$(\mathtt{0x80,0x220  })$}&$2^{-8}$&$2^{-0.58}$&$2^{-8}$&$2^{-16.58}$\\
				\midrule
				\makecell{15\\$(7,4,4)$}&\makecell{$(\mathtt{0x80},\mathtt{0x222})$\\$(\mathtt{0x22},\mathtt{0x8})$}&\makecell{$(\mathtt{0x20},\mathtt{0x80})$\\$(\mathtt{0x20},\mathtt{0x88})$}&$2^{-14}$&$2^{-0.19}$&$2^{-6}$&$2^{-20.19}$\\
				\midrule
				\makecell{15*\\$(5,5,5)$}&\makecell{$(\mathtt{0x8},\mathtt{0x32})$\\$(\mathtt{0x22},\mathtt{0x8})$}&\makecell{$(\mathtt{0x1},\mathtt{0x0})$\\$(\mathtt{0x40000},\mathtt{0x110001})$}&$2^{-8}$&$2^{-0.66}$&$2^{-10}$&$2^{-18.66}$\\
				\midrule
				\makecell{15*\\$(7,3,5)$}&\makecell{$(\mathtt{0x800},\mathtt{0x2220})$\\$(\mathtt{0x220},\mathtt{0x80})$}&\makecell{$(\mathtt{0x11},\mathtt{0x4})$\\$(\mathtt{0x4},\mathtt{0x11})$}&$2^{-14}$&$1$&$2^{-8}$&$2^{-22}$\\
				\midrule
				\makecell{16\\$(7,5,4)$}&\makecell{$(\mathtt{0x400},\mathtt{0x1110})$\\$(\mathtt{0x110},\mathtt{0x40})$}&\makecell{$\mathtt{(0x8, 0x0)}$\\$\mathtt{(0x800008, 0x200000)}$}&$2^{-14}$&$2^{-0.66}$&$2^{-8}$&$2^{-22.66}$\\
				\midrule
				\makecell{16*\\$(7,5,4)$}&\makecell{$(\mathtt{0x400},\mathtt{0x1110})$\\$(\mathtt{0x110},\mathtt{0x40})$}&\makecell{$\mathtt{(0x8, 0x20)}$\\$\mathtt{(0x8, 0x22)}$}&$2^{-14}$&$2^{-1.30}$&$2^{-6}$&$2^{-21.30}$\\
				\midrule
				\makecell{16*\\$(5,6,5)$}&\makecell{$(\mathtt{0x8},\mathtt{0x32})$\\$(\mathtt{0x22},\mathtt{0x8})$}&\makecell{$\mathtt{(0x4, 0x0)}$\\$\mathtt{(0x100000, 0x440004)}$}&$2^{-8}$&$2^{-3.01}$&$2^{-10}$&$2^{-21.01}$\\
				\midrule
				
				\makecell{17\\$(7,6,4)$}&\makecell{$(\mathtt{0x80},\mathtt{0x222})$\\$\mathtt{(0x22,0x8)}$}&\makecell{$\mathtt{(0x0,0x1)}$\\$(\mathtt{0x40000},\mathtt{0x110001})$}&$2^{-14}$&$2^{-0.66}$&$2^{-10}$&$2^{-24.66}$\\
				\midrule
				\makecell{17*\\$(6,6,5)$}&\makecell{$(\mathtt{0x200},\mathtt{0x888})$\\$\mathtt{(0x20,0x8)}$}&\makecell{$\mathtt{(0x4,0x0)}$\\$(\mathtt{0x100000},\mathtt{0x440004})$}&$2^{-12}$&$2^{-2.08}$&$2^{-10}$&$2^{-24.08}$\\
				\midrule
				\makecell{17*\\$(7,4,6)$}&\makecell{$(\mathtt{0x80},\mathtt{0x222})$\\$\mathtt{(0x22,0x8)}$}&\makecell{$\mathtt{(0x400000,0x1)}$\\$(\mathtt{0x40000},\mathtt{0x110001})$}&$2^{-14}$&$1$&$2^{-12}$&$2^{-26}$\\
				\midrule
				\makecell{18*\\$(7,4,7)$}&\makecell{$(\mathtt{0x80},\mathtt{0x222})$\\$\mathtt{(0x22,0x8)}$}&\makecell{$\mathtt{(0x220,0x80)}$\\$(\mathtt{0x8},\mathtt{0x222})$}&$2^{-14}$&$2^{-0.58}$&$2^{-14}$&$2^{-28.58}$\\
				\bottomrule
			\end{tabular}
	\end{threeparttable}
\end{table}

\begin{table}[!h]
	\centering
	\caption{The details of DL trails for $\mathsf{Simon64}$.}{\label{tab:DL-trail-simon64}}
	\renewcommand\tabcolsep{4pt}
	\begin{threeparttable}
			\begin{tabular}{ccccccc}
				\toprule
				Round&$\Delta_I\overset{E_d}{\rightarrow}\Delta$&$\lambda\overset{E_l}{\rightarrow}\lambda_O$&$\textsf{Pro}_{E_d}$&$\mathsf{Cor}_{E_m}$&$\mathsf{Cor}^2_{E_l}$&$\mathsf{Cor}_E$\\
				\midrule
				\makecell{16*\\$(5,6,5)$}&\makecell{$(\mathtt{0x20,0x88 })$\\$(\mathtt{0x88,0x20   })$}&\makecell{$(\mathtt{0x4,0x0 })$\\$(\mathtt{0x1000000,0x4400004  })$}&$2^{-8}$&$2^{-0.59}$&$2^{-10}$&$2^{-18.59}$\\
				\midrule
				\makecell{20\\$(7,7,6)$}&\makecell{$(\mathtt{0x200},\mathtt{0x888})$\\$(\mathtt{0x88},\mathtt{0x20})$}&\makecell{$(\mathtt{0x0},\mathtt{0x4})$\\$(\mathtt{0x1000001},\mathtt{0x4400004})$}&$2^{-14}$&$2^{-0.58}$&$2^{-20}$&$2^{-34.58}$\\
				\midrule
				\makecell{20*\\$(7,7,6)$}&\makecell{$(\mathtt{0x200},\mathtt{0x888})$\\$\mathtt{(0x88,0x20)}$}&\makecell{$\mathtt{(0x20,0x80)}$\\$(\mathtt{0x2},\mathtt{0x80000088})$}&$2^{-14}$&$2^{-7.05}$&$2^{-12}$&$2^{-33.06}$\\\midrule
				\makecell{20*\\$(8,6,6)$}&\makecell{$(\mathtt{0x80},\mathtt{0x8322})$\\$\mathtt{(0x80, 0x22)}$}&\makecell{$\mathtt{(0x80, 0x200)}$\\$(\mathtt{0x8},\mathtt{0x222})$}&$2^{-18}$&$2^{-2.46}$&$2^{-12}$&$2^{-32.46}$\\
				\midrule
				\makecell{20*\\$(9,6,5)$}&\makecell{$(\mathtt{0x800},\mathtt{0x83220})$\\$\mathtt{(0x2220, 0x800)}$}&\makecell{$\mathtt{(0x200, 0x0)}$\\$(\mathtt{0x8},\mathtt{0x222})$}&$2^{-20}$&$2^{-3.16}$&$2^{-10}$&$2^{-33.16}$\\
				\midrule
				\makecell{21*\\$(9,7,5)$}&\makecell{$(\mathtt{0x800},\mathtt{0x83220})$\\$\mathtt{(0x2220, 0x800)}$}&\makecell{$\mathtt{(0x800, 0x0)}$\\$(\mathtt{0x20},\mathtt{0x888})$}&$2^{-20}$&$2^{-6.39}$&$2^{-10}$&$2^{-36.39}$\\
				\midrule
				\makecell{21*\\$(8,5,8)$}&\makecell{$(\mathtt{0x800},\mathtt{0x2220})$\\$\mathtt{(0x800, 0x220)}$}&\makecell{$\mathtt{(0x20, 0x880)}$\\$(\mathtt{0x20},\mathtt{0x888})$}&$2^{-18}$&$2^{-0.91}$&$2^{-18}$&$2^{-36.91}$\\
				\bottomrule
			\end{tabular}
	\end{threeparttable}
\end{table}

\begin{table}[!h]
	\centering
	\caption{The details of DL trails for $\mathsf{Simon96}$.}{\label{tab:DL-trail-simon96}}
	\renewcommand\tabcolsep{4.5pt}
	\begin{threeparttable}
			\begin{tabular}{ccccccc}
				\toprule
				Round&$\Delta_I\overset{E_d}{\rightarrow}\Delta$&$\lambda\overset{E_l}{\rightarrow}\lambda_O$&$\textsf{Pro}_{E_d}$&$\mathsf{Cor}_{E_m}$&$\mathsf{Cor}^2_{E_l}$&$\mathsf{Cor}_E$\\\midrule
				\makecell{23*\\$(9,5,9)$}&\makecell{$(\mathtt{0x10000,0x44400 })$\\$(\mathtt{0x44400,0x10000   })$}&\makecell{$(\mathtt{0x222,0x8 })$\\$(\mathtt{0x8,0x222 })$}&$2^{-20}$&$1$&$2^{-20}$&$2^{-40}$\\
				\midrule
				\makecell{25\\$(10,6,9)$}&\makecell{$(\mathtt{0x11100},\mathtt{0x40400})$\\$(\mathtt{0x11100},\mathtt{0x4000})$}&\makecell{$(\mathtt{0x222},\mathtt{0x8})$\\$(\mathtt{0x8},\mathtt{0x222})$}&$2^{-26}$&$2^{-0.66}$&$2^{-20}$&$2^{-46.66}$\\
				\midrule
				\makecell{25*\\$(9,8,8)$}&\makecell{$(\mathtt{0x800},\mathtt{0x2220})$\\$(\mathtt{0x2220},\mathtt{0x800})$}&\makecell{$(\mathtt{0x20},\mathtt{0x880})$\\$(\mathtt{0x20},\mathtt{0x888})$}&$2^{-20}$&$2^{-6.41}$&$2^{-18}$&$2^{-44.41}$\\
				\midrule
				\makecell{26\\$(11,6,9)$}&\makecell{$(\mathtt{0x101000},\mathtt{0x440400})$\\$(\mathtt{0x44400},\mathtt{0x10000})$}&\makecell{$(\mathtt{0x888},\mathtt{0x20})$\\$(\mathtt{0x20},\mathtt{888})$}&$2^{-30}$&$2^{-0.66}$&$2^{-20}$&$2^{-50.66}$\\
				\midrule
				\makecell{26*\\$(9,9,8)$}&\makecell{$(\mathtt{0x80000},\mathtt{0x222000})$\\$(\mathtt{0x222000},\mathtt{0x80000})$}&\makecell{$(\mathtt{0x1000, 0x4000})$\\$(\mathtt{0x1010, 0x4044})$}&$2^{-20}$&$2^{-8.16}$&$2^{-22}$&$2^{-50.16}$\\
				\bottomrule
			\end{tabular}
	\end{threeparttable}
\end{table}

\begin{table}[!h]
	\centering
	\caption{The details of DL trails for $\mathsf{Simon128}$.}{\label{tab:DL-trail-simon128}}
	\renewcommand\tabcolsep{1.7pt}
	\begin{threeparttable}
			\begin{tabular}{ccccccc}
				\toprule
				Round&$\Delta_I\overset{E_d}{\rightarrow}\Delta$&$\lambda\overset{E_l}{\rightarrow}\lambda_O$&$\textsf{Pro}_{E_d}$&$\mathsf{Cor}_{E_m}$&$\mathsf{Cor}^2_{E_l}$&$\mathsf{Cor}_E$\\\midrule
				\makecell{31\\$(10,9,12)$}&\makecell{$(\mathtt{0x22200000,0x80800000})$\\$(\mathtt{0x22200000,0x8000000})$}&\makecell{$(\mathtt{0x10000,0x440000})$\\$(\mathtt{0x100,0x4044401})$}&$2^{-26}$&$2^{-0.70}$&$2^{-36}$&$2^{-62.70}$\\
				\midrule
				\makecell{31*\\$(13,11,7)$}&\makecell{$(\mathtt{0x800000},\mathtt{0x2220200})$\\$(\mathtt{0x22200},\mathtt{0x8000})$}&\makecell{$(\mathtt{0x100},\mathtt{0x400})$\\$(\mathtt{0x444},\mathtt{0x101})$}&$2^{-38}$&$2^{-5.35}$&$2^{-18}$&$2^{-61.35}$\\\midrule
				\makecell{31*\\$(13,9,9)$}&\makecell{$(\mathtt{0x800000},\mathtt{0x2220200})$\\$(\mathtt{0x22200},\mathtt{0x8000})$}&\makecell{$(\mathtt{0x888},\mathtt{0x20})$\\$(\mathtt{0x20},\mathtt{0x888})$}&$2^{-38}$&$2^{-3.73}$&$2^{-20}$&$2^{-61.73}$\\
				\midrule
				\makecell{32\\$(11,9,12)$}&\makecell{$(\mathtt{0x4040000,0x11010000})$\\$\mathtt{(0x1110000,0x400000)}$}&\makecell{$(\mathtt{0x800,0x22000})$\\$(\mathtt{0x8,0x20222})$}&$2^{-30}$&$2^{-0.70}$&$2^{-36}$&$2^{-66.70}$\\
				\midrule
				\makecell{32*\\$(13,10,9)$}&\makecell{$(\mathtt{0x8000000},\mathtt{0x22202000})$\\$(\mathtt{0x222000},\mathtt{0x80000})$}&\makecell{$(\mathtt{0x4400},\mathtt{0x1000})$\\$(\mathtt{0x1010},\mathtt{0x4044})$}&$2^{-38}$&$2^{-3.61}$&$2^{-24}$&$2^{-65.61}$\\
				\midrule
				\makecell{32*\\$(11,8,13)$}&\makecell{$(\mathtt{0x4040000},\mathtt{0x11010000})$\\$(\mathtt{0x1110000},\mathtt{0x400000})$}&\makecell{$(\mathtt{0x22200},\mathtt{0x800})$\\$(\mathtt{0x8},\mathtt{0x20222})$}&$2^{-30}$&$2^{-0.22}$&$2^{-38}$&$2^{-68.22}$\\
				\bottomrule
			\end{tabular}
	\end{threeparttable}
\end{table}

\begin{table}[!h]
	\centering
	\caption{The details of DL trails for $\mathsf{Simeck32}$.}{\label{tab:DL-trail-simeck32}}
	\renewcommand\tabcolsep{6.5pt}
	\begin{threeparttable}
			\begin{tabular}{ccccccc}
				\toprule
				Round&$\Delta_I\overset{E_d}{\rightarrow}\Delta$&$\lambda\overset{E_l}{\rightarrow}\lambda_O$&$\textsf{Pro}_{E_d}$&$\mathsf{Cor}_{E_m}$&$\mathsf{Cor}^2_{E_l}$&$\mathsf{Cor}_E$\\\midrule
				\makecell{12*\\$(5,2,5)$}&\makecell{$(\mathtt{0x10,0x28 })$\\$(\mathtt{0x28,0x10   })$}&\makecell{$(\mathtt{0x5,0x2 })$\\$(\mathtt{0x2,0x5  })$}&$2^{-8}$&$1$&$2^{-8}$&$2^{-16}$\\
				\midrule
				\makecell{14\\$(5,5,4)$}&\makecell{$(\mathtt{0x2000,0x7400})$\\$(\mathtt{0x400,0x0})$}&\makecell{$(\mathtt{0x100,0x200})$\\$(\mathtt{0x100,0x288})$}&$2^{-10}$&$2^{-0.63}$&$2^{-6}$&$2^{-16.63}$\\
				\midrule
				\makecell{14*\\$(5,6,3)$}&\makecell{$(\mathtt{0x100},\mathtt{0x2a0})$\\$(\mathtt{0x20},\mathtt{0x0})$}&\makecell{$(\mathtt{0x10},\mathtt{0x0})$\\$(\mathtt{0x8},\mathtt{0x14})$}&$2^{-10}$&$2^{-1.99}$&$2^{-4}$&$2^{-15.99}$\\
				\midrule
				\makecell{14*\\$(6,3,5)$}&\makecell{$(\mathtt{0x4},\mathtt{0x800a})$\\$(\mathtt{0x1},\mathtt{0x8000})$}&\makecell{$(\mathtt{0x5100},\mathtt{0x2000})$\\$(\mathtt{0x2000},\mathtt{0x5000})$}&$2^{-12}$&$1$&$2^{-8}$&$2^{-20}$\\
				\bottomrule
			\end{tabular}
	\end{threeparttable}
\end{table}

\begin{table}[!h]
	\centering
	\caption{The details of DL trails for $\mathsf{Simeck48}$.}{\label{tab:DL-trail-simeck48}}
	\renewcommand\tabcolsep{5.5pt}
	\begin{threeparttable}
			\begin{tabular}{ccccccc}
				\toprule
				Round&$\Delta_I\overset{E_d}{\rightarrow}\Delta$&$\lambda\overset{E_l}{\rightarrow}\lambda_O$&$\textsf{Pro}_{E_d}$&$\mathsf{Cor}_{E_m}$&$\mathsf{Cor}^2_{E_l}$&$\mathsf{Cor}_E$\\\midrule
				\makecell{17\\$(6,6,5)$}&\makecell{$(\mathtt{0x80,0x140})$\\$(\mathtt{0x200,0x140})$}&\makecell{$(\mathtt{0x10,0x0})$\\$(\mathtt{0x2,0x15})$}&$2^{-12}$&$2^{-0.37}$&$2^{-10}$&$2^{-22.37}$\\
				\midrule
				\makecell{17*\\$(6,6,5)$}&\makecell{$(\mathtt{0x400, 0xa80})$\\$(\mathtt{0x100, 0x80})$}&\makecell{$(\mathtt{0x28, 0x10})$\\$(\mathtt{0x10, 0x28})$}&$2^{-12}$&$2^{-2.07}$&$2^{-8}$&$2^{-22.07}$\\
				\midrule
				\makecell{17*\\$(5,7,5)$}&\makecell{$(\mathtt{0x800, 0x1500})$\\$(\mathtt{0x100, 0x0})$}&\makecell{$(\mathtt{0x10, 0x0})$\\$(\mathtt{0x2, 0x15})$}&$2^{-10}$&$2^{-1.44}$&$2^{-10}$&$2^{-21.44}$\\
				\midrule
				\makecell{17*\\$(8,4,5)$}&\makecell{$(\mathtt{0x8000, 0x15000})$\\$(\mathtt{0x8000, 0x5000})$}&\makecell{$(\mathtt{0x500, 0x200})$\\$(\mathtt{0x200, 0x500})$}&$2^{-18}$&$1$&$2^{-8}$&$2^{-26}$\\
				\midrule
				\makecell{18\\$(6,6,6)$}&\makecell{$(\mathtt{0x80,0x140})$\\$\mathtt{(0x200,0x140)}$}&\makecell{$(\mathtt{0x10,0x20})$\\$\mathtt{(0x4,0x2a)}$}&$2^{-12}$&$2^{-0.75}$&$2^{-12}$&$2^{-24.75}$\\
				\midrule
				\makecell{18*\\$(8,7,3)$}&\makecell{$(\mathtt{0x80,0x150})$\\$\mathtt{(0x80,0x50)}$}&\makecell{$(\mathtt{0x8, 0x0})$\\$\mathtt{(0x4, 0xa)}$}&$2^{-18}$&$2^{-2.06}$&$2^{-4}$&$2^{-24.06}$\\
				\midrule
				\makecell{18*\\$(7,4,7)$}&\makecell{$(\mathtt{0x8000,0x14000})$\\$\mathtt{(0x54000,0x20000)}$}&\makecell{$(\mathtt{0x2800, 0x1000})$\\$\mathtt{(0x400, 0x2a00)}$}&$2^{-14}$&$1$&$2^{-14}$&$2^{-28}$\\
				\midrule
				\makecell{19*\\$(8,4,7)$}&\makecell{$(\mathtt{0x8000, 0x15000})$\\$(\mathtt{0x8000, 0x5000})$}&\makecell{$(\mathtt{0x500, 0x200})$\\$(\mathtt{0x80, 0x540})$}&$2^{-18}$&$1$&$2^{-14}$&$2^{-32}$\\
				\bottomrule
			\end{tabular}
	\end{threeparttable}
\end{table}

\begin{table}[!h]
	\centering
	\caption{The details of DL trails for $\mathsf{Simeck64}$.}{\label{tab:DL-trail-simeck64}}
	\begin{threeparttable}
			\begin{tabular}{ccccccc}
				\toprule
				Round&$\Delta_I\overset{E_d}{\rightarrow}\Delta$&$\lambda\overset{E_l}{\rightarrow}\lambda_O$&$\textsf{Pro}_{E_d}$&$\mathsf{Cor}_{E_m}$&$\mathsf{Cor}^2_{E_l}$&$\mathsf{Cor}_E$\\\midrule
				\makecell{22\\$(7,7,8)$}&\makecell{$(\mathtt{0x2000,0x5400})$\\$(\mathtt{0x1400,0x800})$}&\makecell{$(\mathtt{0x40,0x280})$\\$(\mathtt{0x40,0x2a0})$}&$2^{-14}$&$2^{-0.44}$&$2^{-18}$&$2^{-32.44}$\\
				\midrule
				\makecell{22*\\$(4,9,9)$}&\makecell{$(\mathtt{0x100, 0x280})$\\$(\mathtt{0x100, 0x80})$}&\makecell{$(\mathtt{0x40, 0x0})$\\$(\mathtt{0x0, 0x44})$}&$2^{-6}$&$2^{-1.04}$&$2^{-22}$&$2^{-29.04}$\\
				\midrule
				\makecell{23\\$(7,7,9)$}&\makecell{$(\mathtt{0x2000,0x5400})$\\$(\mathtt{0x1400,0x800})$}&\makecell{$(\mathtt{0x100,0x0})$\\$(\mathtt{0x0,0x110})$}&$2^{-14}$&$2^{-0.13}$&$2^{-22}$&$2^{-36.13}$\\
				\midrule
				\makecell{23*\\$(4,10,9)$}&\makecell{$(\mathtt{0x100, 0x280})$\\$(\mathtt{0x100, 0x80})$}&\makecell{$(\mathtt{0x40, 0x0})$\\$(\mathtt{0x0, 0x44})$}&$2^{-6}$&$2^{-4.44}$&$2^{-22}$&$2^{-32.44}$\\
				\midrule
				\makecell{23*\\$(4,9,10)$}&\makecell{$(\mathtt{0x100, 0x280})$\\$(\mathtt{0x100, 0x80})$}&\makecell{$(\mathtt{0x20, 0x40})$\\$(\mathtt{0x0, 0x44})$}&$2^{-6}$&$2^{-3.04}$&$2^{-24}$&$2^{-33.04}$\\
				\midrule
				\makecell{24\\$(7,7,10)$}&\makecell{$(\mathtt{0x2000,0x5400})$\\$(\mathtt{0x1400,0x800})$}&\makecell{$(\mathtt{0x100,0x200})$\\$(\mathtt{0x0,0x220})$}&$2^{-14}$&$2^{-0.13}$&$2^{-24}$&$2^{-38.13}$\\
				\midrule
				\makecell{24*\\$(4,9,11)$}&\makecell{$(\mathtt{0x100, 0x280})$\\$(\mathtt{0x100, 0x80})$}&\makecell{$(\mathtt{0x40, 0x0})$\\$(\mathtt{0x20, 0x54})$}&$2^{-6}$&$2^{-1.04}$&$2^{-28}$&$2^{-35.04}$\\
				\midrule
				\makecell{24*\\$(11,6,7)$}&\makecell{$(\mathtt{0x0, 0x880})$\\$(\mathtt{0x28, 0x100})$}&\makecell{$(\mathtt{0x50, 0x20})$\\$(\mathtt{0x8, 0x54})$}&$2^{-26}$&$1$&$2^{-14}$&$2^{-40}$\\
				\midrule
				\makecell{25\\$(7,7,11)$}&\makecell{$(\mathtt{0x2000,0x5400})$\\$(\mathtt{0x1400,0x800})$}&\makecell{$(\mathtt{0x110,0x0})$\\$(\mathtt{0x80,0x140})$}&$2^{-14}$&$2^{-1.04}$&$2^{-26}$&$2^{-41.04}$\\
				\midrule
				\makecell{25*\\$(11,7,7)$}&\makecell{$(\mathtt{0x0,0x880})$\\$\mathtt{(0x280,0x100)}$}&\makecell{$(\mathtt{0x28,0x10})$\\$\mathtt{(0x4,0x2a)}$}&$2^{-26}$&$2^{-1.04}$&$2^{-14}$&$2^{-41.04}$\\
				\midrule
				\makecell{25*\\$(4,10,11)$}&\makecell{$(\mathtt{0x10,0x28})$\\$(\mathtt{0x10,0x8})$}&\makecell{$(\mathtt{0x0,0x4})$\\$(\mathtt{0x4,0x2})$}&$2^{-6}$&$2^{-1.04}$&$2^{-32}$&$2^{-39.04}$\\
				\bottomrule
			\end{tabular}
	\end{threeparttable}
\end{table}

\end{document}